\documentclass[journal, one column]{IEEEtran}
\usepackage{amsmath, amsthm, amsfonts}
\usepackage[english]{babel}
\usepackage[dvips]{graphicx}
\usepackage{graphicx}
\usepackage{tabularx}
\usepackage{enumerate}
\usepackage{mathtools}
\usepackage[ruled]{algorithm2e}
\usepackage{algorithmic}
\usepackage{tikz}
\usepackage{caption}
\usepackage{subcaption}
\newcommand{\pgftextcircled}[1]{
    \setbox0=\hbox{#1}%
    \dimen0\wd0%
    \divide\dimen0 by 2%
    \begin{tikzpicture}[baseline=(a.base)]%
        \useasboundingbox (-\the\dimen0,0pt) rectangle (\the\dimen0,1pt);
        \node[circle,draw,outer sep=0pt,inner sep=0.1ex] (a) {#1};
    \end{tikzpicture}
}

\usepackage{setspace}
\begin{document}
\title{ Diffusion Adaptation Over Clustered Multitask Networks Based on the Affine Projection Algorithm}


\author{{Vinay Chakravarthi Gogineni$^1$, Mrityunjoy Chakraborty$^2$}\\
Department of Electronics and Electrical Communication Engineering\\
Indian Institute of Technology, Kharagpur, INDIA\\
Phone: $+91-3222-283512 \hspace{2em}$ Fax: $+91-3222-255303$\\
E.Mail : $^1\;$vinaychakravarthi@ece.iitkgp.ernet.in, $^2\;$mrityun@ece.iitkgp.ernet.in}

\maketitle
\thispagestyle{empty}

\begin{abstract}
Distributed adaptive networks achieve better estimation performance by exploiting temporal and as well spatial diversity
while consuming few resources.  Recent works have studied the single task distributed estimation problem, in which the nodes
estimate a single optimum parameter vector collaboratively. However, there are many important applications where the multiple
vectors have to estimated simultaneously, in a collaborative manner. This paper presents multi-task diffusion strategies based
on the Affine Projection Algorithm (APA), usage of APA makes the algorithm robust against the correlated input. The performance
analysis of the proposed multi-task diffusion APA algorithm is studied in mean and mean square sense. And also a modified multi-task
diffusion strategy is proposed that improves the performance in terms of convergence rate and steady state EMSE as well. Simulations
are conducted to verify the analytical results.
\end{abstract}
\section{Introduction}
Distributed adaptation over networks has emerged as an attractive and challenging research area with the advent of multi-agent( wireless or wireline) networks. Recent results in the field can be found in $\cite{1}$-$\cite{3}$. In adaptive networks, the interconnected nodes continuously learn and adapt, as well as perform the assigned tasks such as parameter estimation from observations collected by the dispersed agents. Consider a connected network consisting of $N$ nodes observing temporal data arising from different spatial sources with possibly different statistical profiles. The objective is to enable the nodes to estimate a parameter vector of interest, $w_{opt}$ from the observed data. In a centralized approach, the data or local estimates from all nodes would be conveyed to a central processor where they would be fused and the vector of parameters estimated. In order to reduce the requirement of powerful central processor and extensive amount of communications in a traditional centralized solution, a distributed solution is developed relying only on local data exchange and interactions between intermediate neighborhood nodes, while retaining the estimation accuracy of centralized solution. In distributed networks, the individual nodes share the computational burden so that communications are reduced as compared to the centralized network, and power and bandwidth usage are also there by reduced. Due to these merits, distributed estimation has received more attention recently and been widely used in many applications, such as in precision agriculture, environmental monitoring, military surveillance, transportation and instrumentation.
\par
The mode of cooperation that is allowed among the nodes determines the efficiency of any distributed implementation. In incremental mode of cooperation, each node transfers information to its corresponding adjacent node in sequential manner using cyclic pattern of collaboration. This approach reduces communications between nodes and improves the network autonomy as compared the centralized solution. In practical wireless sensor networks, it may be more difficult to establish a cyclic pattern as required in the incremental mode of cooperation as the number of sensor nodes increase. On the other hand, in diffusion mode of cooperation, each node exchanges information with its neighborhood (i.e., the set of all its neighbors including itself), $\mathcal{N}_{k}$ as directed by the network topology. There exist several useful distributed strategies for sequential data processing over networks including consensus strategies $\cite{4}$-$\cite{6}$, incremental strategies $\cite{7}$-$\cite{9}$ and diffusion strategies $\cite{10}$-$\cite{13}$. Diffusion strategies exhibit superior stability and performance over consensus based algorithms \cite{14}.
\par
The existing literature on distributed algorithms shows that most works focus primarily on the case where the nodes estimate a single optimum parameter vector collaboratively. We shall refer to problems of this type as \emph{single-task} problems. However, many problems of interest happen to be multi-task oriented i.e., consider the general situation where there are connected clusters of nodes, and each cluster has a parameter vector to estimate. The estimation still needs to be performed cooperatively across the network because the data across the clusters may be correlated and, therefore, cooperation across clusters can be beneficial. This concept is relevant to the context of distributed estimation and adaptation over networks. Initial investigations along these lines for the traditional diffusion strategy appear in $\cite{15}$-$\cite{19}$. It is well known that in the case of a single adaptive filter, one major drawback of the LMS algorithm is its slow convergence rate for colored input signals and the APA algorithm is a better alternative to LMS is such an environment. For distributed networks, highly correlated inputs also deteriorate the performance of the multi-task diffusion-LMS (multi-task d-LMS) algorithm. In this paper we therefore focus on a new APA-based multi-task distributed learning scheme over networks to obtain a good compromise between convergence performance and computational cost and to analyze their performance in terms of mean-square error and convergence rate.
\section{Network Models and  Multi Task learning}
Consider a network with $N$ nodes deployed over a certain geographical area. At every time instant $n$, every node $k$ has access to time realizations $\{ d_{k}(n), \textbf{u}_{k}(n)\}$  with $d_{k}(n)$ denoting a scalar zero mean reference signal and $\textbf{u}_{k}(n)$ is an  $L \times 1$ regression vector, $\textbf{u}_{k}(n)=[u_{k}(n), u_{k}(n-1), . . . , u_{k}(n-L+1)]^{T} $ with covariance matrix $R_{u, k}=E[\textbf{u}_{k}(n)\textbf{u}^{T}_{k}(n)]$. The data at node $k$ is assumed to be related via the linear measurement model:
\begin{equation}\label{eq2.1}
\begin{split}
d_{k}(n)=\textbf{u}^{T}_{k}(n) \hspace{0.3em} \textbf{w}^{\star}_{k} + \epsilon_{k}(n)
\end{split}
\end{equation}
where $\textbf{w}^{\star}_{k}$ is an unknown optimal parameter vector to be estimated at node $k$ and $\epsilon_{k}(n)$ is an observation noise with variance $\xi^{0}$ which is assumed to be zero mean white noise and also independent of $u_{k}(n)$ for all $k$. Considering the number of parameter vectors to be estimated, which we shall refer to as the number of tasks, the distributed learning problem can be single-task or multi-task oriented. Therefore we distinguish among the following three types of networks, as illustrated by Fig. 1, depending on how the parameter vectors $\textbf{w}^{\star}_{k}$ are related across nodes:
\begin{figure}
        \centering
        \begin{subfigure}[b]{0.3\textwidth}
                \includegraphics[width=\textwidth]{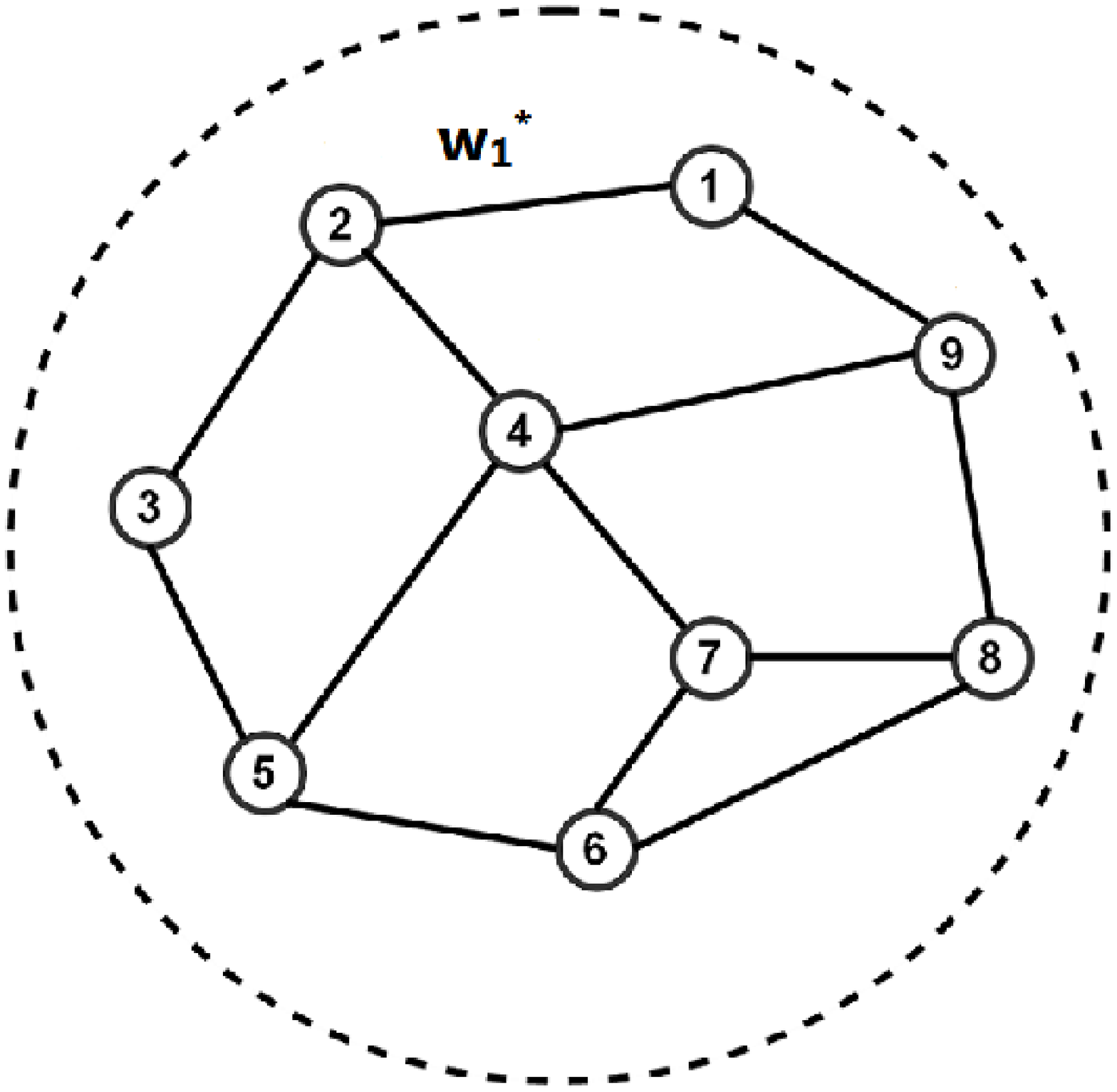}
                \caption{ }
                \label{fig:gull}
        \end{subfigure}%
        ~ 
        \begin{subfigure}[b]{0.3\textwidth}
                \includegraphics[width=\textwidth]{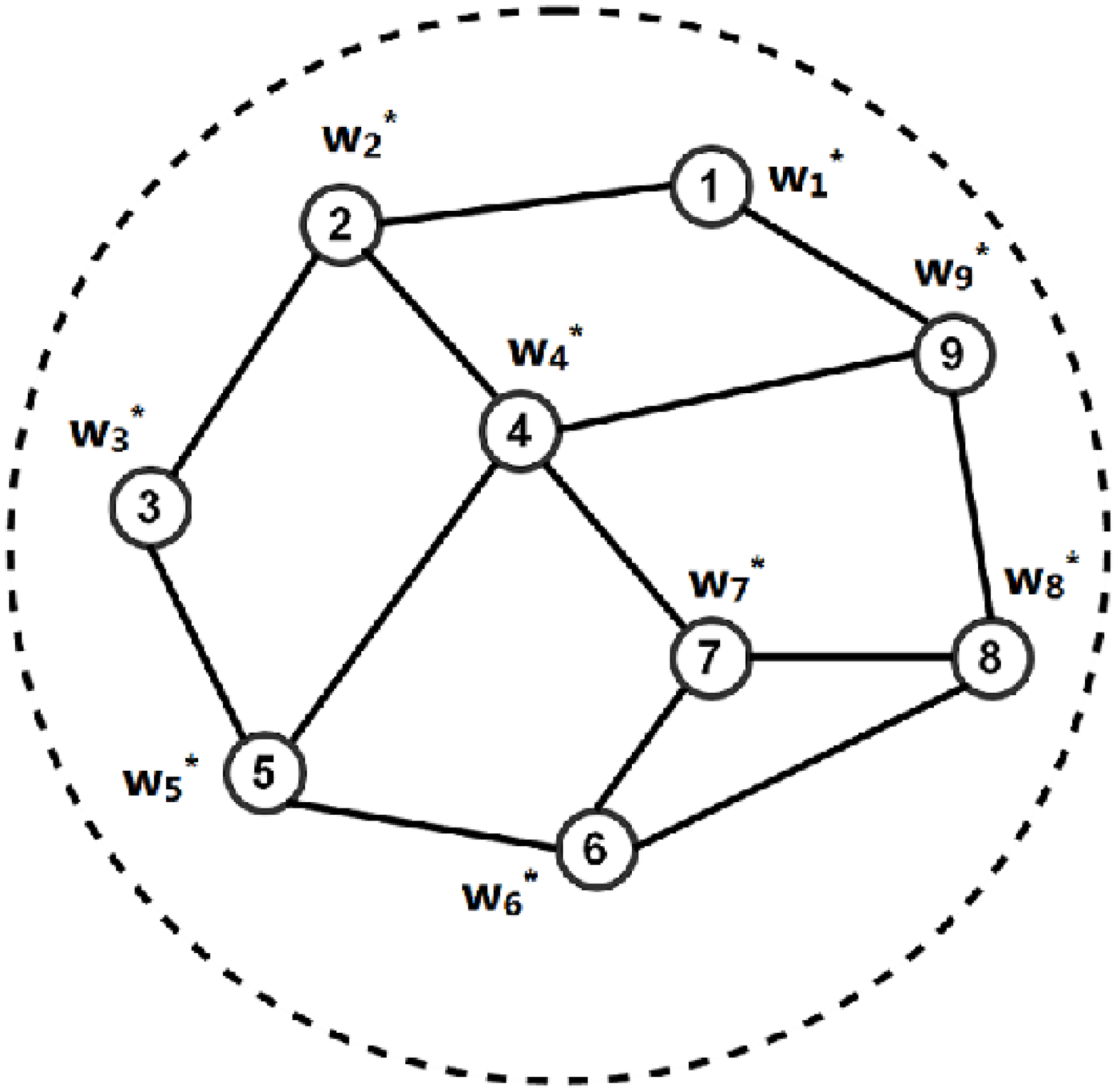}
                \caption{ }
                \label{fig:tiger}
        \end{subfigure}
        ~ 
        \begin{subfigure}[b]{0.3\textwidth}
                \includegraphics[width=\textwidth]{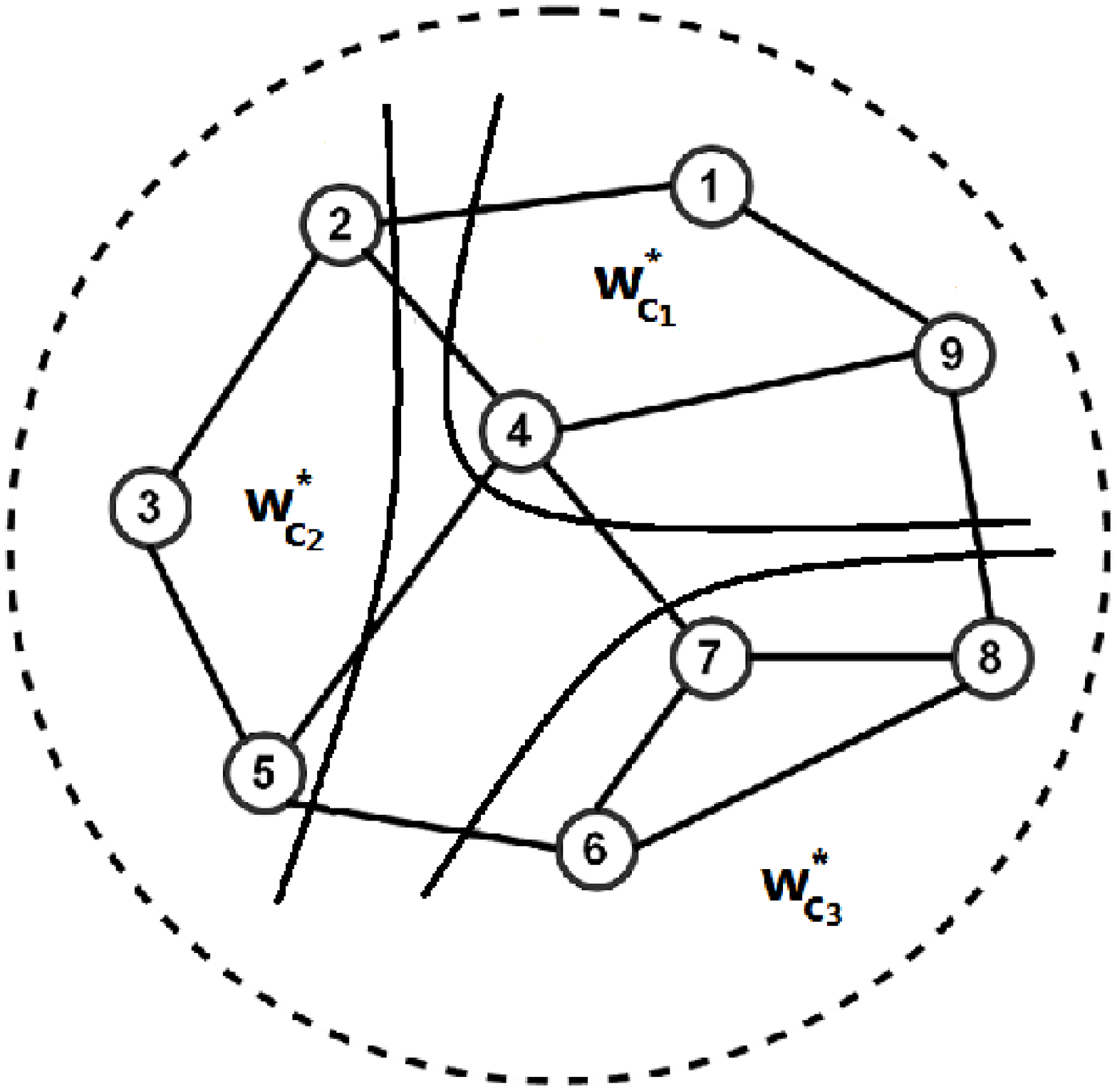}
                \caption{ }
                \label{fig:mouse}
        \end{subfigure}
        \caption{Three types of networks. Through direct links, nodes can communicate with each other in one hop. (a) Single-task Network. (b) Multi-task network. (c) Clustered multi-task network}\label{fig:animals}
\end{figure}
\begin{itemize}
\item \emph{Single-task networks}: All nodes in the network have to estimate the same parameter vector $\textbf{w}^{\star}_{k}$. That is, in this case we have that
\begin{equation}\label{eq2.2}
\begin{split}
\textbf{w}^{\star}_{k}=\textbf{w}^{\star}, \hspace{3em} \forall k \in {1, 2, . . . , N}
\end{split}
\end{equation}
\item \emph{Multi-task networks}: Each node $k$ in the network has to determine its own optimum parameter vector, $\textbf{w}^{\star}_{k}$. However, it is assumed that similarities and relationships exist among the parameters of neighboring nodes, which we denote by writing
\begin{equation}\label{eq2.3}
\begin{split}
\textbf{w}^{\star}_{k} \sim \textbf{w}^{\star}_{l}, \hspace{3em} if \hspace{1em} l \in \mathcal{N}_{k}
\end{split}
\end{equation}
The sign $\sim$ represents a similarity relationship in some sense, and its meaning will become clear soon once we introduce expression (8) and (9) further ahead. There are many situations in practice where the objective parameters are not identical across clusters but have inherent relationships. It is therefore beneficial to exploit these relationships to enhance performance. Here we focus on promoting the similarity of objective parameter vectors via their distance to each other.
\item \emph{Clustered Multi-task Networks}: Nodes are grouped into Q clusters, and there is one task per cluster. The optimum parameter vectors are only constrained to be equal within each cluster. The optimum parameter vectors are only constrained to be equal within each cluster, but similarities between neighboring clusters are allowed to exist, namely,
    \begin{equation}\label{eq2.4}
\begin{split}
\textbf{w}^{\star}_{k} = \textbf{w}^{\star}_{\mathcal{C}_{q}}, \hspace{3em} \text{whenever} \hspace{1em} k \in \mathcal{C}_{q}\hspace{2.8em}\\
\textbf{w}^{\star}_{\mathcal{C}_{p}} \sim \textbf{w}^{\star}_{\mathcal{C}_{q}}, \hspace{3em} \text{if} \hspace{1em} \mathcal{C}_{p}, \mathcal{C}_{q} \hspace{0.5em} \text{are connected}
\end{split}
\end{equation}
where $p$ and $q$ denote two cluster indexes. We say that two clusters $\mathcal{C}_{p}$ and $\mathcal{C}_{p}$ are connected if there exists at least one edge linking a node from one cluster to a node in the other cluster.
\end{itemize}
One can observe that the single-task and multi-task networks are particular cases of the clustered multi-task network. In the case where all nodes are clustered together, the clustered multi-task network reduces to the single-task network. on the other hand, in the case where each cluster only involves one node, the clustered multi-task network becomes a multi-task network. Building on the literature on diffusion strategies for single-task networks, we shall now generalize its usage and analysis for distributed learning over clustered multi-task networks. These results will also be applicable to multi-task networks by setting the number of clusters equal to the number of nodes.
\section{Problem Formulation}
In clustered multitask networks the nodes that are grouped into cluster estimate the same coefficient vector. Thus, consider
the cluster $\mathcal{C}(k)$ to which node $k$ belongs. Under certain settings, in order to provide independence from the input data correlation statistics, we introduce normalized updates with respect to the input regressor at each node $\textbf{u}_{k}(n)$. A local cost function, $J_{k}(\textbf{w}_{\mathcal{C}(k)} )$, is associated with node $k$ and it is assumed that the Hessian matrix of the cost function is positive semi-definite. The local cost function $J_{k}(\textbf{w}_{\mathcal{C}(k)} )$ is defined as
\begin{equation}\label{eq3.1}
\begin{split}
J_{k}(\textbf{w}_{\mathcal{C}(k)}) = E\{ | \frac{\textbf{d}_{k}(n)- \textbf{u}^{T}_{k}(n)\textbf{w}_{\mathcal{C}_{k}}}{\|\textbf{u}_{k}(n)\|}|^{2} \}
\end{split}
\end{equation}
Depending on the application, there may be certain properties among the optimal vectors $\{\textbf{w}^{*}_{\mathcal{C}_{1}}, \hdots,\{\textbf{w}^{*}_{\mathcal{C}_{Q}}\}$. This Mutual information among tasks could be used to improve the estimation accuracy. Among the possible options, a simple yet effective, Euclidian distance based regularizer was enforced in $\cite{18}$. The squared Euclidean distance regularizer is given as
\begin{equation}\label{eq3.2}
\begin{split}
\Delta(\textbf{w}_{\mathcal{C}_{k}}, \textbf{w}_{\mathcal{C}_{l}})= \| \textbf{w}_{\mathcal{C}_{k}} - \textbf{w}_{\mathcal{C}_{l}}\|^{2}
\end{split}
\end{equation}
To estimate the unknown parameter vectors $\{\textbf{w}^{*}_{\mathcal{C}_{1}}, \hdots,\{\textbf{w}^{*}_{\mathcal{C}_{Q}}\}$, it was shown in $\cite{18}$ that the local cost $\eqref{eq3.1}$ and the regularizer $\eqref{eq3.2}$ can be combined at the level of each cluster. This formulation led to the following estimation problem defined in terms of $Q$ Nash equilibrium problems $\cite{20}$, where each cluster $\mathcal{C}_{j}$ estimates $\textbf{w}^{*}_{\mathcal{C}_{j}}$ by minimizing the regularized cost function $J_{\mathcal{C}_{j}}(\textbf{w}_{\mathcal{C}_{j}}, \textbf{w}_{- \mathcal{C}_{j}})$:
\begin{equation}\label{eq3.3}
\begin{split}
(\mathcal{P}_{1})
\begin{cases}
\min\limits_{\textbf{w}_{\mathcal{C}_{j}}} J_{\mathcal{C}_{j}}(\textbf{w}_{\mathcal{C}_{j}}, \textbf{w}_{- \mathcal{C}_{j}}) \\
\text{with} \hspace{0.2em} J_{\mathcal{C}_{j}}(\textbf{w}_{\mathcal{C}_{j}}, \textbf{w}_{- \mathcal{C}_{j}}) = \sum\limits_{k\in \mathcal{C}_{j}} E\{ | \frac{\textbf{d}_{k}(n)- \textbf{u}^{T}_{k}(n)\textbf{w}_{\mathcal{C}_{k}}}{\|\textbf{u}_{k}(n)\|}|^{2} \}  + \eta \sum\limits_{k\in \mathcal{C}_{j}} \sum\limits_{l\in \mathcal{N}_{k} \setminus \mathcal{C}_{j}} \rho_{kl} \| \textbf{w}_{\mathcal{C}_{k}} - \textbf{w}_{\mathcal{C}_{l}}\|^{2}
\end{cases}
\end{split}
\end{equation}
for $j= 1, \hdots, Q$. $\textbf{w}_{\mathcal{C}_{j}}(n)$ is the parameter vector associated with cluster $\mathcal{C}_{j}$, $\eta > 0$ is a regularization parameter, and the symbol $\setminus$ is the set difference. Note that we have kept the notation $\textbf{w}_{\mathcal{C}_{k}}$ in above equation to make the role of the regularization term clearer, even though we have $\textbf{w}_{\mathcal{C}(k)}= \textbf{w}_{\mathcal{C}_{j}}$ for all $k$ in $\mathcal{C}_{j}$. The notation $\textbf{w}_{-\mathcal{C}_{j}}$ denotes the collection of weight vectors estimated by the other clusters, that is, $\textbf{w}_{-\mathcal{C}_{j}}=\{\textbf{w}_{\mathcal{C}_{q}}; q=1, \hdots, Q\}-\{\textbf{w}_{\mathcal{C}_{j}}\}$. The non-negative coefficients $\rho_{kl}$ aim at the adjusting the regularization strength. In $\cite{18}$. The coefficients $\{\rho_{kl}\}$ were chosen to satisfy the conditions:
\begin{equation}\label{eq3.4}
\begin{split}
\sum\limits_{l \in \mathcal{N}_{k}\setminus \mathcal{C}(k)} \rho_{kl} =1,  \text{and} \hspace{0.3em}
\begin{cases}
\rho_{kl} > 0, \text{if} \hspace{0.3em} l \in \mathcal{N}_{k} \setminus \mathcal{C}(k),\\
\rho_{kk} \geq 0,\\
\rho_{kl} = 0, \text{otherwise}
\end{cases}
\end{split}
\end{equation}
We impose $\rho_{kl}=0$ for all $l \notin N_{k}\setminus C(k)$, since nodes belonging to the same cluster estimate the same parameter vector.
\par
The solution for the problem $\mathcal{P}_{1}$ requires that every node in the network should have the access to the statistical moments $\textbf{R}_{u, k}$ and $\textbf{p}_{ud, k}$ over its cluster, however, node $k$ can only be assumed to have direct access to the information from its neighborhood $\mathcal{N}_{k}$, which may include the nodes that are not part of the cluster $\mathcal{C}(k)$ Therefore, to enable a distributed solution that relies only on measured data from neighborhood, as mentioned in $\cite{18}$, $\cite{19}$ the cost function is relaxed into following form:
\begin{equation}\label{eq3.5}
\begin{split}
J_{\mathcal{C}(k)}(\textbf{w}_{k})&= \sum\limits_{l\in \mathcal{N}_{k}\cap\mathcal{C}(k)} c_{lk} E\{ | \frac{\textbf{d}_{l}(n)- \textbf{u}^{T}_{l}(n)\textbf{w}_{k}}{\|\textbf{u}_{k}(n)\|}|^{2} \}
+ \eta \sum\limits_{l\in \mathcal{N}_{k}\setminus\mathcal{C}(k)} \rho_{kl} \|\textbf{w}_{k}-\textbf{w}_{l}\|^{2} \\
& +  \sum\limits_{l\in \mathcal{N}^{-}_{k}\cap\mathcal{C}(k)} b_{lk} \|\textbf{w}_{k}-\textbf{w}^{o}_{l}\|^{2}
\end{split}
\end{equation}
\par
where the coefficients $c_{lk}$ are non-negative and satisfy the conditions:
\begin{equation}\label{eq3.6}
\begin{split}
\sum\limits_{k=1}^{N} c_{lk} =1,  \hspace{0.3em} \text{and} \hspace{0.7em} c_{lk}= 0 \hspace{0.7em} \text{if} \hspace{0.3em} k\notin \mathcal{N}_{l} \cap \mathcal{C}(l)
\end{split}
\end{equation}
and the coefficients $b_{lk}$ are also non-negative.
\par
Following the same line of reasoning from $\cite{10}, \cite{11}$ in the single-task case, and extending the argument to problem $\eqref{eq3.5}$ by using Nash-equilibrium properties $\cite{20}$, and by following same procedure mentioned in $\cite{10}$, $\cite{21}$ the following diffusion strategy of the adapt-then-combine (ATC) for clustered multi-task Normalized LMS (NLMS) is derived in distributed manner:
\begin{equation}\label{eq3.7}
\begin{split}
\begin{cases}
\boldsymbol{\psi}_{k}(n+1) = \textbf{w}_{k}(n)+ \mu \frac{\textbf{u}_{k}(n)}{\|\varepsilon+\textbf{u}_{k}(n)\|^{2}}  [\textbf{d}_{k}(n)- \textbf{u}^{T}_{k}(n)\textbf{w}_{k}(n)] + \mu_{k} \eta \hspace{0.5em} \sum\limits_{l \in \mathcal{N}_{k} \setminus \mathcal{C}(k)}{} \rho_{kl} ( \textbf{w}_{l}(n)-\textbf{w}_{k}(n) )  \\
\textbf{w}_{k}(n+1) = \sum\limits_{l \in \mathcal{N}_{k} \cap \mathcal{C}(k)}{} a_{lk} \hspace{0.2em} \boldsymbol{\psi}_{l}(n+1)   \\
\end{cases}
\end{split}
\end{equation}
By extending the above clustered multi-task diffusion strategy to data-reuse case, we can derive the following Affine projection algorithm (APA) $\cite{22}$ based clustered multi-task diffusion strategy:
\begin{equation}\label{eq3.8}
\begin{split}
\begin{cases}
\boldsymbol{\psi}_{k}(n+1) &= \textbf{w}_{k}(n)+ \mu \textbf{U}^{T}_{k}(n)\left(\varepsilon I + \textbf{U}_{k}(n)\textbf{U}^{T}_{k}(n)\right)^{-1}  [\textbf{d}_{k}(n)- \textbf{u}_{k}(n)\textbf{w}_{k}(n)] \\
& \hspace{2em}+ \mu_{k} \hspace{0.2em} \eta \hspace{1em}\sum\limits_{l \in \mathcal{N}_{k} \setminus \mathcal{C}(k)}{} \rho_{kl} ( \textbf{w}_{l}(n)-\textbf{w}_{k}(n) ) \\
\textbf{w}_{k}(n+1) &= \sum\limits_{l \in \mathcal{N}_{k} \cap \mathcal{C}(k)}{} a_{lk} \hspace{0.2em} \boldsymbol{\psi}_{l}(n+1)   \\
\end{cases}
\end{split}
\end{equation}
where $\eta$ denotes a regularization parameter with small positive value, $\varepsilon$ is employed to avoid the inversion of a rank deficient matrix $\textbf{U}_{k}(n)\textbf{U}^{T}_{k}(n)$ and the input data matrix $\textbf{U}_{k}(n)$, desired response vector $\textbf{d}_{k}(n)$ are given as follows
\begin{equation}\label{eq3.9}
\begin{split}
\textbf{U}_{k}(n)=\left[ \begin{array}{c} \textbf{u}_{k}(n) \\ \textbf{u}_{k}(n-1) \\ \vdots \\ \textbf{u}_{k}(n-P+1)  \end{array} \right] , \hspace{2em}  \textbf{d}_{k}(n)=\left[ \begin{array}{c} d_{k}(n) \\ d_{k}(n-1) \\ \vdots \\ d_{k}(n-P+1)  \end{array} \right]
\end{split}
\end{equation}
The clustered multi-task diffusion APA algorithm is given below:
\begin{algorithm}[h!]
\caption{Diffusion APA for clustered multi-task networks}
\begin{algorithmic}
\item Start $\textbf{w}_{k}(0)=0$ for all $k$, and repeat:
\begin{equation}\label{eq3.10}
\begin{split}
\boldsymbol{\psi}_{k}(n+1)&= \textbf{w}_{k}(n) + \mu_{k} \hspace{0.2em} \textbf{U}^{T}_{k}(n) \left( \varepsilon I + \textbf{U}_{k}(n) \textbf{U}^{T}_{k}(n) \right)^{-1} [\textbf{d}_{k}(n)- \textbf{U}_{k}(n)\textbf{w}_{k}(n)] \\ &\hspace{4.6em} + \hspace{0.2em}\mu_{k} \hspace{0.2em}\eta\hspace{0.2em}\sum\limits_{l \in \mathcal{N}_{k} \setminus \mathcal{C}(k)}{} \rho_{kl} ( \textbf{w}_{l}(n)-\textbf{w}_{k}(n) )\\
\textbf{w}_{k}(n+1)&= \sum\limits_{l \in \mathcal{N}_{k} \cap \mathcal{C}(k)}{} a_{lk} \hspace{0.3em}\boldsymbol{\psi}_{l}(n+1)
\end{split}
\end{equation}
\end{algorithmic}
\end{algorithm}
\par
In a single-task network, there is a single cluster that consists of the entire set of nodes we get $\mathcal{N}_{k} \cap \mathcal{C}(k)=\mathcal{N}_{k}$ and $\mathcal{N}_{k} \setminus \mathcal{C}(k) = \emptyset$ for all $k$, so that the expression $\eqref{eq3.10}$
reduces to the diffusion adaptation strategy $\cite{10}$ as described in algorithm $2$:
\par
\begin{algorithm*}[h!]
\caption{Diffusion APA for single-task networks}
\begin{algorithmic}
\item Start $\textbf{w}_{k}(0)=0$ for all $k$, and repeat:
\begin{equation}\label{eq3.11}
\begin{split}
\boldsymbol{\psi}_{k}(n+1)&= \textbf{w}_{k}(n)+\mu_{k} \hspace{0.2em} \textbf{U}^{T}_{k}(n) \left( \varepsilon I + \textbf{U}_{k}(n) \textbf{U}^{T}_{k}(n) \right)^{-1} [\textbf{d}_{k}(n)- \textbf{U}_{k}(n)\textbf{w}_{k}(n)]\\
\textbf{w}_{k}(n+1)&= \sum\limits_{l \in \mathcal{N}_{k} }{} a_{lk} \hspace{0.3em}\boldsymbol{\psi}_{l}(n+1)
\end{split}
\end{equation}
\end{algorithmic}
\end{algorithm*}
In the case of multi-task network where the size of each cluster is one, we have $\mathcal{N}_{k} \cap \mathcal{C}(k) ={k}$ and $\mathcal{N}_{k} \setminus \mathcal{C}(k) = \mathcal{N}^{-}_{k}$ for all $k$. Then algorithm $1$ degenerates into Algorithm $3$. This is the instantaneous gradient counterpart of $\eqref{eq3.10}$ for each node.
\begin{algorithm}[h!]
\caption{Diffusion APA for multi-task networks}
\begin{algorithmic}
\item Start $\textbf{w}_{k}(0)=0$ for all $k$, and repeat:
\begin{equation}\label{eq3.12}
\begin{split}
\textbf{w}_{k}(n+1)= \textbf{w}_{k}(n) &+ \mu_{k} \hspace{0.2em} \textbf{U}^{T}_{k}(n) \left( \varepsilon I + \textbf{U}_{k}(n) \textbf{U}^{T}_{k}(n) \right)^{-1} [\textbf{d}_{k}(n)- \textbf{U}_{k}(n)\textbf{w}_{k}(n)] \\
& + \hspace{0.2em}\mu_{k} \hspace{0.2em}\eta\hspace{0.2em} \sum\limits_{l \in \mathcal{N}^{-}_{k} }{} \rho_{kl} ( \textbf{w}_{l}(n)-\textbf{w}_{k}(n) )\\
\end{split}
\end{equation}
\end{algorithmic}
\end{algorithm}
\section{Mean-Square Error Performance Analysis}
\subsection{Network Global Model}
The space-time structure of the algorithm leads to challenge in the
performance analysis. To proceed, first, Let us define the global representations as
\begin{equation}\label{eq4.1.1}
\begin{split}
\boldsymbol{\psi}(n)&=\text{col}\{\boldsymbol{\psi}_{1}(n), \boldsymbol{\psi}_{2}(n), \hdots, \boldsymbol{\psi}_{N}(n) \}, \hspace{1em} \textbf{w}(n)=\text{col}\{\textbf{w}_{1}(n), \textbf{w}_{2}(n), \hdots, \textbf{w}_{N}(n) \}\\
\textbf{U}(n)&=\text{diag}\{\textbf{U}_{1}(n), \textbf{U}_{2}(n), \hdots, \textbf{U}_{N}(n) \}, \hspace{1em} \textbf{d}(n)=\text{col}\{\textbf{d}_{1}(n), \textbf{d}_{2}(n), \hdots, \textbf{d}_{N}(n) \}\\
\end{split}
\end{equation}
where $\textbf{U}(n)$ is an $NM \times LN$ block diagonal matrix. The $LN \times LN$ diagonal matrices $\textbf{D}$ and $\boldsymbol{\eta}$ are defined as
\begin{equation}\label{eq4.1.2}
\begin{split}
\textbf{D}&=\text{diag}\{\mu_{1}\textbf{I}_{L}, \mu_{2}\textbf{I}_{L}, \hdots, \mu_{N}\textbf{I}_{L} \}\\
\boldsymbol{\eta}&=\text{diag}\{\eta_{1}\textbf{I}_{L}, \eta_{2}\textbf{I}_{L}, \hdots, \eta_{N}\textbf{I}_{L} \}
\end{split}
\end{equation}
to collect the local step-sizes and regularization parameters. From the linear model of the form $\eqref{eq2.1},$ the global model at network level is obtained as
\begin{equation}\label{eq4.1.3}
\begin{split}
\textbf{d}(n)= \textbf{U}(n) \textbf{w}^{\star} + \textbf{\emph{v}}(n)
\end{split}
\end{equation}
where $\textbf{w}^{\star}(n)$ and $\textbf{v}(n)$ are global optimal weight and noise vectors given as follows
\begin{equation}\label{eq4.1.4}
\begin{split}
\textbf{w}^{\star}(n)&=\text{col}\{\textbf{w}^{\star}_{1}(n), \textbf{w}^{\star}_{2}(n), \hdots, \textbf{w}^{\star}_{N}(n) \}\\
\textbf{\emph{v}}(n)&=\text{col}\{\textbf{\emph{v}}_{1}(n), \textbf{\emph{v}}_{2}(n), \hdots, \textbf{\emph{v}}_{N}(n) \}
\end{split}
\end{equation}
To facilitate analysis, the network topology is assumed to be static (i.e.
$a_{lk}(n) = a_{lk}$). This assumption does not
compromise the algorithm derivation or its operation, and is used for
analysis only. The analysis presented in $\cite{23}$ and $\cite{24}$  serves as the basis for this work. Using the above expressions, the global model of multi-task diffusion APA is therefore formulated as follows:
\begin{equation}\label{eq4.1.5}
\begin{split}
\textbf{w}(n+1)&= \boldsymbol{\mathcal{A}} \hspace{0.2em} \Big[ \textbf{w}(n)+ \textbf{D} \hspace{0.2em} \textbf{U}^{T}(n) \big[\varepsilon \textbf{I} + \textbf{U}(n) \textbf{U}^{T}(n) \big]^{-1} \hspace{0.5em} [\textbf{d}(n)- \textbf{U}(n)\textbf{w}(n)] \hspace{0.2em} - \textbf{D} \hspace{0.1em} \boldsymbol{\eta} \hspace{0.2em} \boldsymbol{\mathcal{Q}} \hspace{0.2em}\textbf{w}(n)\Big]
\end{split}
\end{equation}
where
\begin{equation}\label{eq4.1.6}
\begin{split}
\boldsymbol{\mathcal{A}}&= \textbf{A}^{T} \otimes \textbf{I}_{L} \\
\boldsymbol{\mathcal{Q}}&= \textbf{I}_{LN} - \textbf{P} \otimes \textbf{I}_{L}
\end{split}
\end{equation}
with $\otimes$ denoting the Kronecker product, $\textbf{A}$ is the $N \times N$ symmetric matrix that defines the network topology and $\textbf{P}$ is the $N \times N$ asymmetric matrix that defines regularizer strength among the nodes with $\rho_{kk}=1$ if $\mathcal{N}_{k} \setminus \mathcal{C}(k)$.
Now the objective is to study the performance behavior of the multi-task diffusion APA governed by the form $\eqref{eq4.1.5}$.
\subsection{Mean Error Behavior Analysis}
The global error vector $\textbf{e}(n)$ is related to the local error vectors $\textbf{e}_{k}(n)$ as
\begin{equation}\label{eq4.2.1}
\begin{split}
\textbf{e}(n)&=\text{col}\{\textbf{e}_{1}(n), \textbf{e}_{2}(n), \hdots, \textbf{e}_{N}(n) \}
\end{split}
\end{equation}
By denoting $\widetilde{\textbf{w}}(n)= \textbf{w}^{\star}-\textbf{w}(n)$,
the global weight error vector can be rewritten as
\begin{equation}\label{eqeq4.2.2}
\begin{split}
\textbf{e}(n)=[\textbf{d}(n)- \textbf{U}(n)\textbf{w}(n)] \hspace{0.2em} \Big] = \textbf{U}(n) \widetilde{\textbf{w}}(n) + \textbf{\emph{v}}(n)= \textbf{e}_{a}(n) + \textbf{\emph{v}}(n)
\end{split}
\end{equation}
where
\begin{equation}\label{eq4.2.3}
\begin{split}
\textbf{e}_{a}(n)= \textbf{U}(n) \widetilde{\textbf{w}}(n)
\end{split}
\end{equation}
Using these results the recursive update equation of global weight error vector can be written as
\begin{equation}\label{eq4.2.4}
\begin{split}
\widetilde{\textbf{w}}(n+1)&= \boldsymbol{\mathcal{A}} \hspace{0.2em} \Big[ \widetilde{\textbf{w}}(n)- \textbf{D} \hspace{0.1em} \textbf{U}^{T}(n) \big[\varepsilon \textbf{I} + \textbf{U}(n) \textbf{U}^{T}(n) \big]^{-1} \hspace{0.1em} \textbf{U}(n)\widetilde{\textbf{w}}(n) \hspace{0.1em} - \textbf{D} \hspace{0.1em} \textbf{U}^{T}(n) \big[\varepsilon \textbf{I} + \textbf{U}(n) \textbf{U}^{T}(n) \big]^{-1} \hspace{0.1em} \textbf{\emph{v}}(n) - \hspace{0.1em} \textbf{D} \hspace{0.1em}\boldsymbol{\eta} \hspace{0.2em} \boldsymbol{\mathcal{Q}} \big[\widetilde{\textbf{w}}(n) - \textbf{w}^{\star}\big]\Big] \\
&= \boldsymbol{\mathcal{A}} \hspace{0.2em} \Big[ \textbf{I}_{LN} - \textbf{D} \hspace{0.2em} \textbf{Z}(n) \hspace{0.2em} - \hspace{0.2em} \textbf{D} \hspace{0.2em}\boldsymbol{\eta} \hspace{0.2em} \boldsymbol{\mathcal{Q}} \Big] \hspace{0.2em} \widetilde{\textbf{w}}(n) - \textbf{A}_{I} \hspace{0.1em} \textbf{D} \hspace{0.1em} \textbf{U}^{T}(n) \big[\varepsilon \textbf{I} + \textbf{U}(n) \textbf{U}^{T}(n) \big]^{-1} \hspace{0.1em} \textbf{\emph{v}}(n)  + \boldsymbol{\mathcal{A}} \hspace{0.2em} \textbf{D} \hspace{0.2em}\boldsymbol{\eta} \hspace{0.2em} \boldsymbol{\mathcal{Q}} \hspace{0.2em} \textbf{w}^{\star} \\
\end{split}
\end{equation}
where $\textbf{Z}(n) = \textbf{U}^{T}(n) \big[\varepsilon \textbf{I} + \textbf{U}(n) \textbf{U}^{T}(n) \big]^{-1} \hspace{0.1em} \textbf{U}(n)$.
Taking the expectation $E[\cdot]$ on both sides, and using the statistical independence between $\textbf{w}_{k}(n)$ and $\textbf{U}_{k}(n)$ (i.e., "independence assumption"), and recalling that $\textbf{v}_{k}(n)$ is zero-mean
i.i.d and also independent of $\textbf{U}_{k}(n)$ and thus of $\textbf{w}_{k}(n)$ we can write
\begin{equation}\label{eq4.2.5}
\begin{split}
E\big[\widetilde{\textbf{w}}(n+1)\big]= \boldsymbol{\mathcal{A}} \bigg[ \textbf{I}_{LN} - \textbf{D} \hspace{0.2em} E\big[ \textbf{Z}(n) \big]\hspace{0.2em} - \hspace{0.2em} \textbf{D} \hspace{0.2em}\boldsymbol{\eta} \hspace{0.2em} \boldsymbol{\mathcal{Q}} \bigg] E[\widetilde{\textbf{w}}(n)] +  \boldsymbol{\mathcal{A}} \hspace{0.2em} \textbf{D} \hspace{0.2em}\boldsymbol{\eta} \hspace{0.2em} \boldsymbol{\mathcal{Q}} \hspace{0.2em} \textbf{w}^{\star}
\end{split}
\end{equation}
Then, for any initial condition, in order to guarantee the stability of the multi-task diffusion APA strategy in the mean sense, the step size $\mu_{k}$ has to be chosen to satisfy
\begin{equation}\label{eq4.2.6}
\begin{split}
 \lambda_{max} \Big(\hspace{0.1em} \boldsymbol{\mathcal{A}} \big[\textbf{I}_{LN}-\textbf{D} \hspace{0.2em} \overline{\textbf{Z}}-\textbf{D} \hspace{0.2em}\boldsymbol{\eta} \hspace{0.2em} \boldsymbol{\mathcal{Q}} \big] \hspace{0.1em} \Big) < 1
\end{split}
\end{equation}
where $\overline{\textbf{Z}}=E[\textbf{Z}(n)]$, and $\lambda_{max}(\cdot)$ denotes the maximum eigen value of its argument matrix.
Therefore, using the norm inequalities and recalling the fact that the combining matrix $\textbf{A}$ is a left stochastic matrix ( i.e., block maximum norm is equal to one), we have
\begin{equation}\label{eq4.2.8}
\begin{split}
\| \boldsymbol{\mathcal{A}} \big[\textbf{I}_{LN}-\textbf{D}\hspace{0.2em} \overline{\textbf{Z}}-\textbf{D} \hspace{0.2em}\boldsymbol{\eta} \hspace{0.2em} \boldsymbol{\mathcal{Q}} \big] \|_{b, \infty} &\leq \| \big[\textbf{I}_{LN}-\textbf{D}\hspace{0.2em} \overline{\textbf{Z}}-\textbf{D} \hspace{0.2em}\boldsymbol{\eta} \hspace{0.2em} \boldsymbol{\mathcal{Q}} \big] \|_{b, \infty}\\
& \leq \| \big[\textbf{I}_{LN}-\textbf{D}\hspace{0.2em} \overline{\textbf{Z}}-\textbf{D} \hspace{0.2em}\boldsymbol{\eta} \hspace{0.2em}  + \textbf{D} \hspace{0.2em}\boldsymbol{\eta} \hspace{0.2em} \big( \textbf{P} \otimes \textbf{I}_{L}\big)  \big] \|_{b, \infty}
\end{split}
\end{equation}
Let $A$ be the an $L \times L$ matrix, then from Gershgorin circle theorem, we have:
\begin{equation}
 |\lambda - a_{i, i} | \leq \sum\limits_{j \neq i} |a_{i, j}|
\end{equation}
Therefore, using the above result, and recalling the fact that $\textbf{P}$ is a right stochastic matrix, a sufficient condition for $\eqref{eq4.2.8}$ to hold is to choose $\mu$ such that
\begin{equation}\label{eq4.2.9}
\begin{split}
0 < \mu_{k} < \frac{2}{\text{max}_{k}\{\lambda_{max}( \hspace{0.2em} \overline{\textbf{Z}}_{k})\}+2\eta}
\end{split}
\end{equation}
where $\overline{\textbf{Z}}_{k}=E\Big[\textbf{U}_{k}^{T}(n) \big[\varepsilon \textbf{I} + \textbf{U}_{k}(n) \textbf{U}_{k}^{T}(n) \big]^{-1} \hspace{0.5em} \textbf{U}_{k}(n) \Big]$.
Above result clearly shows that the mean stability limit of the clustered multi-task diffusion APA is lower than the diffusion
APA due to the presence of $\eta$.
\par
In steady-state i.e., as $n \rightarrow \infty $  the asymptotic mean bias is given by
\begin{equation}\label{eq4.2.10}
\begin{split}
\lim\limits_{n \to \infty}E\big[\widetilde{\textbf{w}}(n)\big]= \Bigg[ \textbf{I}_{LN} -  \boldsymbol{\mathcal{A}} \hspace{0.2em} \bigg[ \textbf{I}_{LN} - \textbf{D} \hspace{0.2em} \overline{\textbf{Z}} \hspace{0.2em} - \hspace{0.2em} \textbf{D} \hspace{0.2em} \boldsymbol{\eta} \hspace{0.2em} \boldsymbol{\mathcal{Q}} \bigg]  \Bigg]^{-1}   \boldsymbol{\mathcal{A}} \hspace{0.2em} \textbf{D }\hspace{0.2em} \boldsymbol{\eta} \hspace{0.2em}  \boldsymbol{\mathcal{Q}} \hspace{0.2em} \textbf{w}^{\star}
\end{split}
\end{equation}
\subsection{Mean-Square Error Behavior Analysis}
The recursive update equation of weight error vector can also be rewritten as follows:
\begin{equation}\label{eq4.3.1}
\begin{split}
\widetilde{\textbf{w}}(n+1)&= \boldsymbol{\mathcal{G}}(n) \hspace{0.3em} \widetilde{\textbf{w}}(n) - \boldsymbol{\mathcal{A}} \hspace{0.2em} \textbf{D}\hspace{0.1em}\textbf{U}^{T}(n) \big[\varepsilon \textbf{I} + \textbf{U}(n) \textbf{U}^{T}(n) \big]^{-1} \textbf{\emph{v}}(n) + \textbf{r}
\end{split}
\end{equation}
where
\begin{equation}\label{eq4.3.2}
\begin{split}
\boldsymbol{\mathcal{G}}(n)&= \boldsymbol{\mathcal{A}} \hspace{0.2em} \Big[\textbf{I}_{LN} -  \textbf{D} \hspace{0.2em} \textbf{Z}(n) - \textbf{D}\hspace{0.2em}\boldsymbol{\eta} \hspace{0.1em}\boldsymbol{\mathcal{Q}}\Big]\\
\textbf{r}&=\boldsymbol{\mathcal{A}} \hspace{0.2em}\textbf{D} \hspace{0.2em} \boldsymbol{\eta} \hspace{0.2em}\boldsymbol{\mathcal{Q}} \hspace{0.2em} \textbf{w}^{\star}
\end{split}
\end{equation}
Using the standard independent assumption between $\textbf{U}_{k}(n)$ and $\textbf{w}_{k}(n)$ and $E[\textbf{v}(n)]=0$, the mean square of the weight error vector $\widetilde{\textbf{w}}(n+1)$, weighted by any positive semi-definite matrix $\boldsymbol{\Sigma}$ that we are free to choose, satisfies the following relation:
\begin{equation}\label{eq4.3.3}
\begin{split}
E\|\widetilde{\textbf{w}}(n+1)\|_{\boldsymbol{\Sigma}}^{2}&= E\|\widetilde{\textbf{w}}(n)\|_{E \boldsymbol{\Sigma}^{'}}^{2}
+ E\big[\textbf{\emph{v}}^{T}(n) \hspace{0.1em}\textbf{Y}^{\boldsymbol{\Sigma}}(n) \hspace{0.1em} \textbf{\emph{v}}(n) \big]+ E\big[\widetilde{\textbf{w}}^{T}(n)\big] \hspace{0.2em} E\big[\boldsymbol{\mathcal{G}}^{T}(n) \big]\hspace{0.1em}\boldsymbol{\Sigma} \hspace{0.2em} \textbf{r} + \textbf{r}^{T} \hspace{0.2em} \boldsymbol{\Sigma} \hspace{0.2em} E\big[\boldsymbol{\mathcal{G}}(n)\big] \hspace{0.2em} E\big[\widetilde{\textbf{w}}(n)\big] + \|\textbf{r}\|_{\boldsymbol{\Sigma}}^{2}
\end{split}
\end{equation}
where
\begin{equation}\label{eq4.3.4}
\begin{split}
E\boldsymbol{\Sigma}^{'}&= E \Big[\boldsymbol{\mathcal{G}}^{T}(n) \hspace{0.2em} \boldsymbol{\Sigma} \hspace{0.2em} \boldsymbol{\mathcal{G}}(n) \Big]\\
&=\hspace{0.2em}\boldsymbol{\mathcal{A}}^{T} \hspace{0.1em} \boldsymbol{\Sigma} \hspace{0.2em} \boldsymbol{\mathcal{A}} - E \big[\textbf{Z}(n)] \hspace{0.2em} \textbf{D} \hspace{0.1em} \boldsymbol{\mathcal{A}}^{T} \hspace{0.1em} \boldsymbol{\Sigma} \hspace{0.1em} \boldsymbol{\mathcal{A}} - \boldsymbol{\mathcal{A}}^{T} \hspace{0.1em} \boldsymbol{\Sigma} \hspace{0.1em} \boldsymbol{\mathcal{A}} \hspace{0.1em} \textbf{D} \hspace{0.1em} E \big[\textbf{Z}(n) \big] - \boldsymbol{\mathcal{A}}^{T} \hspace{0.1em} \boldsymbol{\Sigma} \hspace{0.1em} \textbf{S} - \textbf{S}^{T}\hspace{0.1em} \boldsymbol{\Sigma} \hspace{0.1em} \boldsymbol{\mathcal{A}} \hspace{0.1em}\\
&\hspace{0.2em} + E \big[\textbf{Z}(n) \big] \textbf{D} \hspace{0.1em} \boldsymbol{\mathcal{A}}^{T}_{I} \hspace{0.1em} \boldsymbol{\Sigma} \hspace{0.1em} \textbf{S} + \textbf{S}^{T} \hspace{0.1em} \boldsymbol{\Sigma} \hspace{0.1em} \boldsymbol{\mathcal{A}} \hspace{0.1em} \textbf{D} \hspace{0.1em} E \big[ \textbf{Z}(n) \big] + E \big[\textbf{U}^{T}(n)\hspace{0.1em} \textbf{Y}^{\boldsymbol{\Sigma}}\hspace{0.1em} \textbf{U}(n) \big] + \textbf{S}^{T} \hspace{0.1em} \boldsymbol{\Sigma} \hspace{0.2em} \textbf{S}
\end{split}
\end{equation}
and
\begin{equation}\label{eq4.3.5}
\begin{split}
\textbf{Y}^{\boldsymbol{\Sigma}}&=\big[\varepsilon \textbf{I} + \textbf{U}(n) \textbf{U}^{T}(n) \big]^{-1}\textbf{U}(n) \hspace{0.2em} \textbf{D} \hspace{0.2em} \boldsymbol{\mathcal{A}}^{T} \hspace{0.2em} \boldsymbol{\Sigma} \hspace{0.2em} \boldsymbol{\mathcal{A}} \hspace{0.2em} \textbf{D} \hspace{0.2em}    \textbf{U}^{T}(n) \big[\varepsilon \textbf{I} + \textbf{U}(n) \textbf{U}^{T}(n) \big]^{-1}\\
\textbf{S}&= \boldsymbol{\mathcal{A}} \hspace{0.2em} \textbf{D} \hspace{0.2em} \boldsymbol{\eta} \hspace{0.2em} \boldsymbol{\mathcal{Q}}
\end{split}
\end{equation}
In order to study the behavior of the multi-task diffusion APA algorithm, the
following moments in $\eqref{eq4.3.3}$ and $\eqref{eq4.3.4}$ must be evaluated:
\begin{equation}\label{eq4.3.6}
\begin{split}
&E\Big[\textbf{Z}(n) \hspace{0.2em}\textbf{D} \hspace{0.2em} \boldsymbol{\mathcal{A}}^{T} \hspace{0.2em} \boldsymbol{\Sigma} \hspace{0.2em} \boldsymbol{\mathcal{A}} \hspace{0.2em} \textbf{D} \hspace{0.2em}  \textbf{Z}(n) \Big] \\
&E\Big[\textbf{\emph{v}}^{T}(n) \big[\varepsilon \textbf{I} + \textbf{U}(n) \textbf{U}^{T}(n) \big]^{-1} \textbf{U}(n) \hspace{0.2em} \textbf{D} \hspace{0.2em} \boldsymbol{\mathcal{A}}^{T} \hspace{0.2em} \boldsymbol{\Sigma} \hspace{0.2em} \boldsymbol{\mathcal{A}} \hspace{0.2em} \textbf{D} \hspace{0.2em} \textbf{U}^{T}(n)\big[\varepsilon \textbf{I} + \textbf{U}(n) \textbf{U}^{T}(n) \big]^{-1} \textbf{\emph{v}}(n) \Big]
\end{split}
\end{equation}
To extract the matrix $\boldsymbol{\Sigma}$ from the expectation terms, a weighted variance relation is introduced by using $L^{2}N^{2} \times 1$ column vectors:
\begin{equation}\label{eq4.3.7}
\begin{split}
\boldsymbol{\sigma}= \text{bvec}\{\boldsymbol{\Sigma}\} \hspace{2em} \text{and} \hspace{2em} \boldsymbol{\sigma^{'}}= \text{bvec}\{E\boldsymbol{\Sigma}^{'}\}
\end{split}
\end{equation}
where $\text{bvec}\{\cdot\}$ denotes the block vector operator. In addition, $\text{bvec}\{\cdot\}$ is also used to recover the original matrix $\Sigma$ from $\boldsymbol{\sigma}$. One property of the $\text{bvec}\{\cdot\}$ operator when working with
the block Kronecker product $\cite{26}$ is used in this work, namely,
\begin{equation}\label{eq4.3.8}
\begin{split}
\text{bvec}\{\textbf{Q} \boldsymbol{\Sigma} \textbf{P}\}= ( \textbf{P}^{T} \otimes_{b} \textbf{Q} ) \hspace{0.2em} \boldsymbol{\sigma}
\end{split}
\end{equation}
where $\textbf{P} \otimes_{b} \textbf{Q}$ denotes the block Kronecker product $\cite{25}$, $\cite{26}$ of two block matrices.
\par
Using $\eqref{eq4.3.8}$ to $\eqref{eq4.3.4}$ after block vectorization, the following terms on the
right side of $\eqref{eq4.3.4}$ are given by
\begin{equation}\label{eq4.3.9}
\begin{split}
\text{bvec} \Big\{ \boldsymbol{\mathcal{A}}^{T} \hspace{0.2em} \boldsymbol{\Sigma} \hspace{0.2em} \boldsymbol{\mathcal{A}} \Big \}&= \Big( \boldsymbol{\mathcal{A}}^{T} \otimes_{b} \boldsymbol{\mathcal{A}}^{T} \Big) \hspace{0.2em} \boldsymbol{\sigma}
\end{split}
\end{equation}
\begin{equation}\label{eq4.3.10}
\begin{split}
\text{bvec}\Big\{E\big[\textbf{Z}(n)\big] \hspace{0.2em} \textbf{D} \hspace{0.2em} \boldsymbol{\mathcal{A}}^{T} \boldsymbol{\Sigma} \hspace{0.2em} \boldsymbol{\mathcal{A}} \Big\}&=  \Big( \textbf{I}_{LN} \otimes_{b} E\big[\textbf{Z}(n)\big] \Big) \hspace{0.2em} \Big( \textbf{I}_{LN} \otimes_{b} \textbf{D} \Big) \hspace{0.2em} \Big( \boldsymbol{\mathcal{A}}^{T} \otimes_{b} \boldsymbol{\mathcal{A}}^{T} \Big) \hspace{0.2em} \boldsymbol{\sigma}
\end{split}
\end{equation}
\begin{equation}\label{eq4.3.11}
\begin{split}
\text{bvec} \Big\{ \boldsymbol{\mathcal{A}}^{T} \boldsymbol{\Sigma} \hspace{0.2em} \boldsymbol{\mathcal{A}} \hspace{0.2em} \textbf{D} \hspace{0.2em} E[\textbf{Z}(n)] \Big \}&= \Big( E\big[\textbf{Z}(n) \big] \otimes_{b} \textbf{I}_{LN} \Big)  \Big( \textbf{D}  \otimes_{b} \textbf{I}_{LN} \Big) \Big( \boldsymbol{\mathcal{A}}^{T}  \otimes_{b} \boldsymbol{\mathcal{A}}^{T} \Big ) \hspace{0.2em} \boldsymbol{\sigma}
\end{split}
\end{equation}
\begin{equation}\label{eq4.3.12}
\begin{split}
\text{bvec}\Big\{\boldsymbol{\mathcal{A}}^{T}  \hspace{0.2em} \boldsymbol{\Sigma} \hspace{0.2em} \textbf{S} \Big\}&=\text{bvec}\Big\{ \boldsymbol{\mathcal{A}}^{T}  \hspace{0.2em} \boldsymbol{\Sigma} \hspace{0.2em} \boldsymbol{\mathcal{A}}  \hspace{0.2em} \textbf{D} \hspace{0.2em} \boldsymbol{\eta} \hspace{0.2em} \boldsymbol{\mathcal{Q}} \Big\}\\
&= \Big( \boldsymbol{\mathcal{Q}}^{T} \otimes_{b} \textbf{I}_{LN} \Big ) \Big( \boldsymbol{\eta}  \otimes_{b} \textbf{I}_{LN} \Big ) \Big( \textbf{D}  \otimes_{b} \textbf{I}_{LN} \Big) \Big( \boldsymbol{\mathcal{A}}^{T} \otimes_{b} \boldsymbol{\mathcal{A}}^{T} \Big) \hspace{0.2em}\boldsymbol{\sigma}
\end{split}
\end{equation}
\begin{equation}\label{eq4.3.13}
\begin{split}
\text{bvec}\Big\{\textbf{S}^{T} \hspace{0.2em}\boldsymbol{\Sigma} \hspace{0.2em} \boldsymbol{\mathcal{A}} \Big\}&=\text{bvec}\Big\{\boldsymbol{\mathcal{Q}}^{T} \hspace{0.2em} \boldsymbol{\eta}  \hspace{0.2em} \textbf{D} \hspace{0.2em} \boldsymbol{\mathcal{A}}^{T}  \hspace{0.2em} \boldsymbol{\Sigma} \hspace{0.2em} \boldsymbol{\mathcal{A}}  \Big\}\\
&= \Big( \textbf{I}_{LN} \otimes_{b}  \boldsymbol{\mathcal{Q}}^{T} \Big) \Big( \textbf{I}_{LN} \otimes_{b} \boldsymbol{\eta}  \Big) \Big( \textbf{I}_{LN} \otimes_{b} \textbf{D}  \Big) \Big( \boldsymbol{\mathcal{A}}^{T} \otimes_{b}\boldsymbol{\mathcal{A}}^{T} \Big) \hspace{0.2em} \boldsymbol{\sigma}
\end{split}
\end{equation}
\begin{equation}\label{eq4.3.14}
\begin{split}
\text{bvec}\Big\{E\big[\textbf{Z}(n) \big] \hspace{0.2em} \textbf{D} \hspace{0.2em}  \boldsymbol{\mathcal{A}}^{T}  \boldsymbol{\Sigma} \hspace{0.2em} \textbf{S} \Big\}&=\text{bvec}\Big\{ E\big[\textbf{Z}(n)\big] \hspace{0.2em} \textbf{D} \hspace{0.2em} \boldsymbol{\mathcal{A}}^{T}  \Sigma \hspace{0.2em} \boldsymbol{\mathcal{A}} \hspace{0.2em} \textbf{D} \hspace{0.2em} \boldsymbol{\eta} \hspace{0.2em} \boldsymbol{\mathcal{Q}}  \Big\}\\
&= \Big( \textbf{I}_{LN} \otimes_{b} E\big[\textbf{Z}(n)\big]   \Big) \Big( \boldsymbol{\mathcal{Q}}^{T}  \otimes_{b}\textbf{I}_{LN}  \Big) \Big(\boldsymbol{\eta}  \otimes_{b} \textbf{I}_{LN}\Big) \Big(\textbf{D} \otimes_{b} \textbf{D} \Big) \Big( \boldsymbol{\mathcal{A}}^{T}  \otimes_{b} \boldsymbol{\mathcal{A}} \Big) \hspace{0.2em} \boldsymbol{\sigma}
\end{split}
\end{equation}

\begin{equation}\label{eq4.3.15}
\begin{split}
\text{bvec}\Big\{\textbf{S}^{T} \boldsymbol{\Sigma} \hspace{0.2em} \boldsymbol{\mathcal{A}} \hspace{0.2em} \textbf{D} \hspace{0.2em} E\big[\textbf{Z}(n) \big]  \Big\}&=\text{bvec}\Big\{ \boldsymbol{\mathcal{Q}}^{T} \hspace{0.2em} \boldsymbol{\eta}  \hspace{0.1em} \textbf{D} \hspace{0.1em} \boldsymbol{\mathcal{A}}^{T}  \hspace{0.2em} \boldsymbol{\Sigma} \hspace{0.2em} \boldsymbol{\mathcal{A}} \hspace{0.2em} \textbf{D} \hspace{0.2em} E\big[\textbf{Z}(n) \big] \Big\}\\
&= \Big( E[\textbf{Z}(n)]  \odot \textbf{I}_{LN} \Big) \Big(\textbf{I}_{LN} \otimes_{b}  \boldsymbol{\mathcal{Q}}^{T} \Big) \Big(\textbf{I}_{LN} \otimes_{b} \boldsymbol{\eta}  \Big) \Big( \textbf{D}  \otimes_{b} \textbf{D}  \Big) \Big( \boldsymbol{\mathcal{A}}^{T} \otimes_{b} \boldsymbol{\mathcal{A}}^{T} \Big) \hspace{0.2em} \boldsymbol{\sigma}
\end{split}
\end{equation}

\begin{equation}\label{eq4.3.16}
\begin{split}
\text{bvec}\Big\{ \textbf{U}^{T}(n) \textbf{Y}^{\boldsymbol{\Sigma}} \hspace{0.2em} \textbf{U}(n) \Big\}&=\text{bvec}\Big\{E\big[\textbf{Z}(n) \hspace{0.2em} \textbf{D} \hspace{0.2em} \boldsymbol{\mathcal{A}}^{T}  \hspace{0.2em} \boldsymbol{\Sigma} \hspace{0.2em} \boldsymbol{\mathcal{A}} \hspace{0.2em} \textbf{D} \hspace{0.2em} \textbf{Z}(n) \big] \Big\}\\
&= E\big[ \textbf{Z}(n) \otimes_{b} \textbf{Z}(n)\big] \Big(\textbf{D} \otimes_{b} \textbf{D}  \Big)  \Big( \boldsymbol{\mathcal{A}}^{T}  \otimes_{b} \boldsymbol{\mathcal{A}}^{T} \Big) \hspace{0.2em} \boldsymbol{\sigma}
\end{split}
\end{equation}

\begin{equation}\label{eq4.3.17}
\begin{split}
\text{bvec}\Big\{ \textbf{S}^{T} \Sigma \hspace{0.2em} \textbf{S} \Big\}&=\text{bvec}\Big\{\hspace{0.2em} \boldsymbol{\mathcal{Q}}^{T}   \boldsymbol{\eta}  \hspace{0.2em} \textbf{D}  \hspace{0.2em} \boldsymbol{\mathcal{A}}^{T}  \boldsymbol{\Sigma} \hspace{0.2em} \boldsymbol{\mathcal{A}}  \hspace{0.2em} \textbf{D} \hspace{0.2em} \boldsymbol{\eta} \hspace{0.2em} \boldsymbol{\mathcal{Q}} \Big\}\\
&= \Big(  \boldsymbol{\mathcal{Q}}^{T} \otimes_{b} \boldsymbol{\mathcal{Q}}^{T}   \Big) \Big(\boldsymbol{\eta}  \otimes_{b} \boldsymbol{\eta}  \Big) \Big(\textbf{D} \otimes_{b} \textbf{D}  \Big)  \Big( \boldsymbol{\mathcal{A}}^{T}  \otimes_{b} \boldsymbol{\mathcal{A}}^{T} \Big) \hspace{0.2em} \boldsymbol{\sigma}
\end{split}
\end{equation}
Therefore, a linear relation between the corresponding vectors $\{\boldsymbol{\sigma},\boldsymbol{\sigma}^{'}\}$is formulated by
\begin{equation}\label{eq4.3.18}
\begin{split}
\boldsymbol{\sigma}^{'}= \textbf{F} \boldsymbol{\sigma}
\end{split}
\end{equation}
where $\textbf{F}$ is an $L^{2}N^{2} \times L^{2}N^{2}$ matrix and given by
\begin{equation}\label{eq4.3.19}
\begin{split}
\textbf{F}&=\left[\begin{array}{l}  \textbf{I}_{LN}- \big(\textbf{I}_{LN} \otimes_{b} \overline{\textbf{Z}} \big)  \big(\textbf{I}_{LN} \otimes_{b} \textbf{D} \big)  - \big(  \overline{\textbf{Z}}  \otimes_{b} \textbf{I}_{LN}\big)  \big(\textbf{D}  \otimes_{b} \textbf{I}_{LN}\big) \\
-\big( \boldsymbol{\mathcal{Q}}^{T} \otimes_{b} \textbf{I}_{LN}\big) \big(\boldsymbol{\eta}  \otimes_{b} \textbf{I}_{LN}\big) \big(\textbf{D}  \otimes_{b} \textbf{I}_{LN}\big) -  \big(\textbf{I}_{LN} \otimes_{b}  \boldsymbol{\mathcal{Q}}^{T} \big) \big(\textbf{I}_{LN} \otimes_{b} \boldsymbol{\eta}  \big) \big(\textbf{I}_{LN} \otimes_{b} \textbf{D}  \big) \\
+ \big( \textbf{I}_{LN} \otimes_{b} \overline{\textbf{Z}}  \big) \big( \boldsymbol{\mathcal{Q}}^{T}  \otimes_{b} \textbf{I}_{LN}  \big) \big(\boldsymbol{\eta}  \otimes_{b} \textbf{I}_{LN}\big) \big(\textbf{D}  \otimes_{b} \textbf{D} \big)\\  + \big( \overline{\textbf{Z}}  \otimes_{b} \textbf{I}_{LN}\big) \big(\textbf{I}_{LN} \otimes_{b}  \boldsymbol{\mathcal{Q}}^{T} \big) \big(\textbf{I}_{LN} \otimes_{b} \boldsymbol{\eta}  \big) \big(\textbf{D}  \otimes_{b} \textbf{D}  \big) +  \boldsymbol{\Pi} \big(\textbf{D}  \otimes_{b} \textbf{D}  \big)\\ +  \big(  \boldsymbol{\mathcal{Q}}^{T}  \otimes_{b} \boldsymbol{\mathcal{Q}}^{T} \big) \big(\boldsymbol{\eta}  \otimes_{b} \boldsymbol{\eta}  \big) \big(\textbf{D}  \otimes_{b} \textbf{D}  \big)   \end{array} \right] \big( \boldsymbol{\mathcal{A}}^{T} \otimes_{b} \boldsymbol{\mathcal{A}}^{T} \big)
\end{split}
\end{equation}
where $\boldsymbol{\Pi}=E\Big[\textbf{Z} (n) \otimes_{b} \textbf{Z}(n)  \Big]$
Let $\Lambda_{v} = E[\textbf{\emph{v}}(n)\textbf{\emph{v}}^{T}(n)]$ denote a $NM \times NM$  diagonal matrix, whose entries
are the noise variances $\sigma^{2}_{v,k}$ for $k = 1,2, \cdots ,N$ and given by
\begin{equation}\label{eq4.3.20}
\begin{split}
\Lambda_{v}=\text{diag}\{\sigma^{2}_{v, 1}\textbf{I}_{M}, \sigma^{2}_{v, 2}\textbf{I}_{M}, \cdots, \sigma^{2}_{v, N}\textbf{I}_{M}\}
\end{split}
\end{equation}
Using the independence assumption of noise signals, the term $E\big[\textbf{\emph{v}}^{T}(n) \hspace{0.1em}\textbf{Y}^{\boldsymbol{\Sigma}}(n) \hspace{0.1em} \textbf{\emph{v}}(n) \big]$ can be written as
\begin{equation}\label{eq4.3.21}
\begin{split}
E\big[\textbf{\emph{v}}^{T}(n) \hspace{0.2em} \textbf{Y}^{\boldsymbol{\Sigma}}(n) \hspace{0.2em} \textbf{\emph{v}}(n) \big]&= Tr\Big(\boldsymbol{\mathcal{A}}  \hspace{0.2em} \textbf{D} \hspace{0.2em} E[\boldsymbol{\Phi}]\hspace{0.2em} \textbf{D} \hspace{0.2em} \boldsymbol{\mathcal{A}}^{T}  \boldsymbol{\Sigma} \Big)\\
&= \boldsymbol{\gamma}^{T} \hspace{0.2em} \boldsymbol{\sigma}
\end{split}
\end{equation}
where $\boldsymbol{\Phi}= \textbf{U}^{T}(n)\big[\varepsilon \textbf{I} + \textbf{U}(n) \textbf{U}^{T}(n) \big]^{-1} \boldsymbol{\Lambda}_{v}(n) \big[\varepsilon \textbf{I} + \textbf{U}(n) \textbf{U}^{T}(n) \big]^{-1} \textbf{U}(n) $
and
\begin{equation}\label{eq4.3.22}
\begin{split}
\boldsymbol{\gamma}&= \text{vec}\big\{ \boldsymbol{\mathcal{A}}  \hspace{0.2em}\textbf{D} E[\boldsymbol{\Phi}] \hspace{0.2em} \textbf{D}^{T}\hspace{0.2em} \boldsymbol{\mathcal{A}}^{T}  \big\}\\
&= \big( \boldsymbol{\mathcal{A}}   \otimes \boldsymbol{\mathcal{A}}   \big)  \big(\textbf{D}  \otimes \textbf{D}  \big) \text{vec}\big\{ E[\textbf{W}^{T} \boldsymbol{\Lambda}_{v} \textbf{W} ]  \big\} \\
&= \big( \boldsymbol{\mathcal{A}}   \otimes \boldsymbol{\mathcal{A}} \big)  \big(\textbf{D}  \otimes \textbf{D}  \big)  E\big[\big(\textbf{W}^{T} \otimes \textbf{W}^{T} \big)\big] \hspace{0.2em} \boldsymbol{\gamma}_{v}
\end{split}
\end{equation}
with $\textbf{W}=\big[\varepsilon \textbf{I} + \textbf{U}(n) \textbf{U}^{T}(n) \big]^{-1}   \textbf{U}(n)$ and $\boldsymbol{\gamma}_{v}=\text{vec}\{\boldsymbol{\Lambda}_{v}\}$.

Finally, let us define the $\emph{\textbf{f}}\big(\textbf{r}, E[\widetilde{\textbf{w}}(n)], \boldsymbol{\sigma}\big)$ as the last three terms on the right hand side of the $\eqref{eq4.3.3}$, i.e,
\begin{equation}\label{eq4.3.23}
\begin{split}
\emph{\textbf{f}}\big(\textbf{r}, E[\widetilde{\textbf{w}}(n)], \boldsymbol{\sigma}\big)&=  \|\textbf{r}\|_{\boldsymbol{\Sigma}}^{2} + E\big[\widetilde{\textbf{w}}^{T}(n)\big] \hspace{0.2em} E\big[\boldsymbol{\mathcal{G}}^{T}(n) \big] \hspace{0.2em} \boldsymbol{\Sigma} \hspace{0.2em} \textbf{r} + \textbf{r}^{T} \hspace{0.2em} \boldsymbol{\Sigma} \hspace{0.2em} E\big[\boldsymbol{\mathcal{G}}(n)\big]\hspace{0.1em} E\big[\widetilde{\textbf{w}}(n)\big] \\
\end{split}
\end{equation}
Each term can be evaluated as follows. Now, let us consider the term $E\|\textbf{r} \|_{\boldsymbol{\Sigma}}^{2} $, that can be written
\begin{equation}
\begin{split}
\|\textbf{r} \|_{\boldsymbol{\Sigma}}^{2}&= \bigg(\text{bvec}\Big\{ \boldsymbol{\mathcal{A}} \hspace{0.2em} \textbf{D} \hspace{0.2em} \boldsymbol{\eta} \hspace{0.2em} \boldsymbol{\mathcal{Q}} \hspace{0.2em} \textbf{w}^{\star} \big(\textbf{w}^{\star}\big)^{T}  \boldsymbol{\mathcal{Q}}^{T}  \boldsymbol{\eta}  \hspace{0.2em} \textbf{D} \hspace{0.2em}  \boldsymbol{\mathcal{A}}^{T} \Big\} \bigg)^{T} \hspace{0.2em} \boldsymbol{\sigma}\\
&= \textbf{r}^{T}_{b} \boldsymbol{\sigma}
\end{split}
\end{equation}
where
\begin{equation}
\begin{split}
\textbf{r}_{b}&= E\Big( \boldsymbol{\mathcal{A}}  \otimes_{b} \boldsymbol{\mathcal{A}}  \Big) \big( \textbf{D} \otimes_{b} \textbf{D} \big) \big( \boldsymbol{\eta} \otimes_{b} \boldsymbol{\eta} \big) \Big( \boldsymbol{\mathcal{Q}}  \otimes_{b} \boldsymbol{\mathcal{Q}}  \Big)  \text{bvec}\Big\{ \textbf{w}^{\star} \big(\textbf{w}^{\star}\big)^{T} \Big\}\\
\end{split}
\end{equation}
Consider the second term $E\Big[ \widetilde{\textbf{w}}^{T}(n) \hspace{0.2em}  \boldsymbol{\mathcal{G}}^{T}(n)  \hspace{0.2em}\boldsymbol{\Sigma} \hspace{0.3em}\textbf{r}  \Big]$ that can be simplified as follows:
\begin{equation}
\begin{split}
E\Big[ \widetilde{\textbf{w}}^{T}(n) \hspace{0.2em}  \boldsymbol{\mathcal{G}}^{T}(n)  \hspace{0.1em}\boldsymbol{\Sigma} \hspace{0.3em}\textbf{r}  \Big] &=
Tr\Big(E\Big[  \textbf{r}  \hspace{0.2em} \widetilde{\textbf{w}}^{T}(n) \hspace{0.3em} \boldsymbol{\mathcal{G}}^{T}(n) \Big] \hspace{0.2em} \boldsymbol{\Sigma} \Big)\\
&=\boldsymbol{\alpha}^{T}_{1}(n) \hspace{0.2em} \boldsymbol{\sigma}
\end{split}
\end{equation}
where
\begin{equation}
\begin{split}
\boldsymbol{\alpha}_{1}(n)&=  \Big( \boldsymbol{\mathcal{A}}  \otimes_{b} \boldsymbol{\mathcal{A}}  \Big) \left[\begin{array}{l}  \big( \textbf{I}_{LN} \otimes_{b}  \textbf{D}  \big)  \big( \textbf{I}_{LN} \otimes_{b} \boldsymbol{\eta} \big) \Big( \textbf{I}_{LN} \otimes_{b} \boldsymbol{\mathcal{Q}}   \Big) \\
-  \big( \textbf{D}  \otimes_{b} \textbf{D}  \big)  \big( \textbf{I}_{LN} \otimes_{b} \boldsymbol{\eta} \big) \Big( \textbf{I}_{LN} \otimes_{b} \boldsymbol{\mathcal{Q}}   \Big) \big( \overline{\textbf{Z}} \otimes_{b} \textbf{I}_{LN} \big)  \\
-  \big( \textbf{D}  \otimes_{b} \textbf{D}  \big) \big( \boldsymbol{\eta} \otimes_{b} \boldsymbol{\eta} \big)  \Big( \boldsymbol{\mathcal{Q}}  \otimes_{b} \boldsymbol{\mathcal{Q}}  \Big) \end{array}\right] \text{bvec}\Big\{\textbf{w}^{\star} E[\widetilde{\textbf{w}}^{T}(n) ] \Big\}\\
\end{split}
\end{equation}
In the same way, third term  $E\Big[ \textbf{r}^{T}  \hspace{0.3em}\boldsymbol{\Sigma}\hspace{0.2em} \boldsymbol{\mathcal{G}}(n) \hspace{0.2em} \widetilde{\textbf{w}}(n)\Big]$ can be written as follows:
\begin{equation}
\begin{split}
E\Big[ \textbf{r}^{T}  \hspace{0.3em} \boldsymbol{\Sigma} \hspace{0.2em} \boldsymbol{\mathcal{G}}(n) \hspace{0.2em} \widetilde{\textbf{w}}(n) \Big] &=
Tr\Big(E\Big[ \boldsymbol{\mathcal{G}}(n) \hspace{0.2em} \widetilde{\textbf{w}}(n) \hspace{0.2em}\textbf{r}^{T}  \Big] \hspace{0.2em} \boldsymbol{\Sigma} \Big)\\
&=\boldsymbol{\alpha}^{T}_{2}(n)  \hspace{0.2em} \boldsymbol{\sigma}
\end{split}
\end{equation}
where
\begin{equation}
\begin{split}
\boldsymbol{\alpha}_{2}(n)&=  \Big( \boldsymbol{\mathcal{A}}  \otimes_{b} \boldsymbol{\mathcal{A}}  \Big) \left[\begin{array}{l}  \big( \textbf{D}  \otimes_{b} \textbf{I}_{LN} \big)  \big( \boldsymbol{\eta} \otimes_{b} \textbf{I}_{LN} \big) \Big(   \boldsymbol{\mathcal{Q}}  \otimes_{b} \textbf{I}_{LN} \Big) \\
-  \big( \textbf{D}  \otimes_{b} \textbf{D}  \big)  \big( \boldsymbol{\eta} \otimes_{b} \textbf{I}_{LN} \big) \Big(  \boldsymbol{\mathcal{Q}}  \otimes_{b} \textbf{I}_{LN} \Big)  \big( \textbf{I}_{LN} \otimes_{b} \overline{\textbf{Z}} \big)  \\
- \big( \textbf{D}  \otimes_{b} \textbf{D}  \big) \big( \boldsymbol{\eta} \otimes_{b} \boldsymbol{\eta} \big)  \Big( \boldsymbol{\mathcal{Q}}  \otimes_{b} \boldsymbol{\mathcal{Q}}  \Big) \end{array}\right] \text{bvec}\Big\{ E[\widetilde{\textbf{w}}(n)] \big(\textbf{w}^{\star} \big)^{T}  \Big\}\\
\end{split}
\end{equation}
Therefore, the mean-square behavior of the multi-task diffusion APA algorithm is summarized
as follows:
\begin{equation}\label{eq4.3.24}
\begin{split}
E\|\widetilde{\textbf{w}}(n+1)\|_{\boldsymbol{\sigma}}^{2}&= E\|\widetilde{\textbf{w}}(n)\|_{\textbf{F}\boldsymbol{\sigma}}^{2}
+ \boldsymbol{\gamma}^{T} \hspace{0.1em} \boldsymbol{\sigma}+\emph{\textbf{f}}\big(\textbf{r}, E[\widetilde{\textbf{w}}(n)], \boldsymbol{\sigma}\big)\\
&= E\|\widetilde{\textbf{w}}(n)\|_{\textbf{F}\boldsymbol{\sigma}}^{2}
+ \boldsymbol{\gamma}^{T} \hspace{0.1em} \boldsymbol{\sigma}+ \Big(\textbf{r}^{T}_{b} + \boldsymbol{\alpha}^{T}_{1}(n) +\boldsymbol{\alpha}^{T}_{2}(n)  \Big)\hspace{0.2em} \boldsymbol{\sigma}
\end{split}
\end{equation}
Therefore, the multi-task diffusion strategy presented in $\eqref{eq3.10}$ is mean square stable if the matrix $\textbf{F}$ is stable. Iterating the recursion $\eqref{eq4.3.24}$ starting from $n=0$, we get
\begin{equation}\label{eq4.3.25}
\begin{split}
E\|\widetilde{\textbf{w}}(n+1)\|_{\boldsymbol{\sigma}}^{2}&= E\|\widetilde{\textbf{w}}(0)\|_{\textbf{F}^{n+1}\boldsymbol{\sigma}}^{2}
+ \boldsymbol{\gamma}^{T} \hspace{0.1em} \sum\limits_{i=0}^{n}\textbf{F}^{i}\boldsymbol{\sigma}+\sum\limits_{i=0}^{n}\emph{\textbf{f}}\big(\textbf{r}, E[\widetilde{\textbf{w}}(n-i)], \textbf{F}^{i} \boldsymbol{\sigma}\big)
\end{split}
\end{equation}
with initial condition $\widetilde{\textbf{w}}(0)=\textbf{w}^{\star} - \textbf{w}(0)$. If the matrix $\textbf{F}$ is stable then the first and second terms in the above equation converge to a finite value as $n \rightarrow \infty $. Now, let us consider the third term on the RHS of the $\eqref{eq4.3.25}$. We know that $E[\widetilde{\textbf{w}}(n)]$ is uniformly bounded because $\eqref{eq4.2.5}$ is a BIBO stable recursion with bounded driving term $ \hspace{0.2em} \boldsymbol{\mathcal{A}} \hspace{0.2em}\textbf{D} \hspace{0.2em} \boldsymbol{\eta} \hspace{0.2em} \boldsymbol{\mathcal{Q}} \hspace{0.2em} \textbf{w}^{\star}$. Therefore, from $\eqref{eq4.3.23}$  $\emph{\textbf{f}}\big(\textbf{r}, E[\widetilde{\textbf{w}}(n-i)], \textbf{F}^{i} \boldsymbol{\sigma}\big)$ can be written as
\begin{equation}\label{eq4.3.26}
\begin{split}
\emph{\textbf{f}}\big(\textbf{r}, E[\widetilde{\textbf{w}}(n-i)], \textbf{F}^{i} \boldsymbol{\sigma}\big)&= \Big(\textbf{r}^{T}_{b} + \boldsymbol{\alpha}^{T}_{1}(n-i) +\boldsymbol{\alpha}^{T}_{2}(n-i)  \Big) \hspace{0.2em} \textbf{F}^{i} \hspace{0.2em} \boldsymbol{\sigma}
\end{split}
\end{equation}
Provided that $\textbf{F}$ is stable and there exist a matrix norm, denoted by $\|\cdot\|_{p}$ such that $\|\textbf{F}\|_{p}=c_{p} < 1$. Applying this norm to $\emph{\textbf{f}}$ and using the matrix norms and triangular inequality, we can write $ \| \emph{\textbf{f}}\big(\textbf{r}, E[\widetilde{\textbf{w}}(n-i)], \textbf{F}^{i} \boldsymbol{\sigma}\big)  \| \leq \emph{v} \hspace{0.1em} c^{i}_{p}$, given $\emph{v}$ is a small positive constant. Therefore $E\|\widetilde{\textbf{w}}(n+1)\|_{\boldsymbol{\sigma}}^{2}$ converges to a bounded value as $n \rightarrow \infty$, and the algorithm is said to be mean square stable.
\par
By selecting $\boldsymbol{\Sigma}=\frac{1}{N} \textbf{I}_{LN}$ we can relate $E\|\widetilde{\textbf{w}}(n+1)\|_{\boldsymbol{\sigma}}^{2}$ and $E\|\widetilde{\textbf{w}}(n)\|_{\boldsymbol{\sigma}}^{2}$ as follows:
\begin{equation}\label{eq4.3.27}
\begin{split}
E\|\widetilde{\textbf{w}}(n+1)\|_{\boldsymbol{\sigma}}^{2}&=E\|\widetilde{\textbf{w}}(n)\|_{\boldsymbol{\sigma}}^{2} + \boldsymbol{\gamma}^{T} \textbf{F}^{n} \boldsymbol{\sigma} - E\|\widetilde{\textbf{w}}(0)\|_{\big(I_{(LN)^{2}} - \textbf{F} \hspace{0.1em}\big)\textbf{F}^{n} \boldsymbol{\sigma}}^{2} + \sum\limits_{i=0}^{n}\emph{\textbf{f}}\big(\textbf{r}, E[\widetilde{\textbf{w}}(n-i)], \textbf{F}^{i}\boldsymbol{\sigma}\big) \\
& \hspace{2em} - \sum\limits_{i=0}^{n-1}\emph{\textbf{f}}\big(\textbf{r}, E[\widetilde{\textbf{w}}(n-1-i)], \textbf{F}^{i}\boldsymbol{\sigma}\big)
\end{split}
\end{equation}
we can rewrite the last two terms in the above equation as,
\begin{equation}\label{eq4.3.28}
\begin{split}
&\sum\limits_{i=0}^{n}\emph{\textbf{f}}\big(\textbf{r}, E[\widetilde{\textbf{w}}(n-i)], \textbf{F}^{i}\boldsymbol{\sigma}\big)
 - \sum\limits_{i=0}^{n-1}\emph{\textbf{f}}\big(\textbf{r}, E[\widetilde{\textbf{w}}(n-1-i)], \textbf{F}^{i}\boldsymbol{\sigma}\big) =  \textbf{r}_{b}^{T} \hspace{0.2em} \textbf{F}^{n} \hspace{0.2em} \boldsymbol{\sigma} + \left[   \boldsymbol{\alpha}_{1}^{T}(n) + \boldsymbol{\alpha}_{2}^{T}(n)   + \Gamma(n)  \right] \hspace{0.2em}  \boldsymbol{\sigma}
\end{split}
\end{equation}
where
\begin{equation}\label{eq4.3.29}
\begin{split}
\boldsymbol{\Gamma} (n) =   \sum\limits_{i=1}^{n} \Big( \boldsymbol{\alpha}_{1}^{T}(n-i) +  \boldsymbol{\alpha}_{2}^{T}(n-i) \Big) \textbf{F}^{i} \hspace{0.2em} \boldsymbol{\sigma} - \sum\limits_{i=0}^{n-1} \Big( \boldsymbol{\alpha}_{1}^{T}(n-1-i) +  \boldsymbol{\alpha}_{2}^{T}(n-1-i) \Big) \textbf{F}^{i} \hspace{0.2em} \boldsymbol{\sigma}
\end{split}
\end{equation}
Therefore, the recursion presented in $\eqref{eq4.3.24}$ can be rewritten as,
\begin{equation}\label{eq4.3.30}
\begin{split}
E\|\widetilde{\textbf{w}}(n+1)\|_{\boldsymbol{\sigma}}^{2}&= E\|\widetilde{\textbf{w}}(n)\|_{\boldsymbol{\sigma}}^{2} + \boldsymbol{\gamma}^{T} \textbf{F}^{n} \boldsymbol{\sigma} - E\|\widetilde{\textbf{w}}(0)\|_{\big(I_{(LN)^{2}} - \textbf{F} \hspace{0.1em}\big)\textbf{F}^{n} \boldsymbol{\sigma}}^{2} +\textbf{r}_{b}^{T} \textbf{F}^{n} \hspace{0.1em} \boldsymbol{\sigma} + \left[   \boldsymbol{\alpha}_{1}^{T}(n) + \boldsymbol{\alpha}_{2}^{T}(n)   + \Gamma(n)  \right] \\
\boldsymbol{\Gamma} (n+1)&= \boldsymbol{\Gamma} (n) \textbf{F} + \Big[  \big[ \boldsymbol{\alpha}_{1}^{T}(n) + \boldsymbol{\alpha}_{2}^{T}(n)  \big]\hspace{0.2em} [\textbf{F}- \textbf{I}_{(LN)^{2}}]     \Big]
\end{split}
\end{equation}
with $\boldsymbol{\Gamma} (0)= 0_{1 \times (LN)^{2}}$.
\par
Steady-state MSD of the multi-task diffusion APA strategy is given as follows
\begin{equation}\label{eq4.3.31}
\begin{split}
\lim\limits_{n \to \infty}E\|\widetilde{\textbf{w}}(n)\|_{\big(I_{(LN)^{2}}-\textbf{F}\big) \boldsymbol{\sigma}}^{2}&= \boldsymbol{\gamma}^{T} \boldsymbol{\sigma} + \emph{\textbf{f}}\big(\textbf{r}, E[\widetilde{\textbf{w}}(\infty)], \boldsymbol{\sigma}\big)
\end{split}
\end{equation}
\subsection{New Approach to Improve the Performance of Clustered multi-task diffusion APA}
Clustered multi-task diffusion strategy presented in $\eqref{eq3.10}$ has mainly $2$ drawbacks
\begin{itemize}
\item At time instance $n$, assume that the node $l$ exhibiting poor performance over the node $k$. The multi-task diffusion strategy forces the node $k$ to learn from node $l$ during the adaptation step where $l \in N_{k}\setminus C_{k}$. This affects the performance in transient state.
\item For all $l \in N_{k} \setminus C_{k}$  we have $\textbf{w}_{k}^{*} \simeq \textbf{w}_{l}^{*}$ i.e, only the underlying system is same. However, the multitask diffusion strategy forces the node $k$ to learn from node $l$ even in the steady state. This hampers the steady state performance of the algorithm.
\end{itemize}
To address these problems a control variable called similarity measure, $\delta_{kl}(n)$ is introduced to control the regularizer term in
the multi-task diffusion strategy. At each time instance $n$, node $k$ has access to its neighborhood filter coefficient vectors. Since the node is learning from its neighborhood filter coefficient vectors, it is reasonable to check the similarity among the filter coefficient vectors. The similarity measure is calculated as follows
\begin{equation}\label{eq4.4.1}
\begin{split}
\delta_{kl}(n)= \frac{1}{2}\bigg[1 + sign\Big( \hspace{0.2em} \sigma^{2}_{k}(n) - \sigma^{2}_{kl}(n) \hspace{0.2em} \Big) \bigg]
\end{split}
\end{equation}
where $\sigma^{2}_{k}(n)$ and $\sigma^{2}_{lk}(n)$ are estimated error variances and can be calculated as
\begin{equation}\label{eq4.4.2}
\begin{split}
\sigma^{2}_{k}(n)&= \lambda \hspace{0.2em} \sigma^{2}_{k}(n-1) + (1-\lambda) \Big[ d_{k}(n) - \textbf{u}^{T}_{k}(n) \textbf{w}_{k}(n) \Big]^{2}\\
\sigma^{2}_{kl}(n)&= \lambda \hspace{0.2em} \sigma^{2}_{kl}(n-1) + (1-\lambda) \Big[ d_{k}(n) - \textbf{u}^{T}_{k}(n) \textbf{w}_{l}(n) \Big]^{2}  \hspace{1em} \text{for} \hspace{1em} l \in \mathcal{N}_{k}\setminus \mathcal{C}(k)
\end{split}
\end{equation}
and $\lambda$ is a positive constant with $\lambda \in [0, 1]$.
\par
To explain, suppose that at index $n$, the node $l$ performs better than node $k$, i.e., $\sigma^{2}_{kl}(n) < \sigma^{2}_{k}(n)$. Then for node $k$ the similarity measure $\delta_{kl}(n)=\frac{1}{2}\Big[1 + sign\big( \hspace{0.2em}\sigma^{2}_{k}(n) - \sigma^{2}_{kl}(n)  \hspace{0.2em} \big) \Big]= 1$, which implies that node $k$ would learn the weight information from node $l$ by adding the difference of their current weight vectors, i.e., $[\textbf{w}_{l}(n) - \textbf{w}_{k}(n)]$ as a correction term to its weight update. On the other hand, suppose the node $l$ does not perform better than node $k$, i.e., $\sigma^{2}_{kl}(n) > \sigma^{2}_{k}(n)$. Then for node $k$ the similarity measure $\delta_{kl}(n)=\frac{1}{2}\Big[ 1 + sign \big( \hspace{0.2em} \sigma^{2}_{k}(n) - \sigma^{2}_{kl}(n)   \hspace{0.2em} \big) \Big]= 0$, which implies that node $k$ would neglect the weight vector $\textbf{w}_{l}(n)$. Thus improves the convergence rate and steady state performance over the multi-task diffusion strategy presented in $\eqref{eq3.10}$.
\par
Therefore, by taking the similarity measure, $\delta_{kl}(n)$ into account the modified clustered multi-task diffusion APA is given below
\begin{equation}\label{eq4.4.3}
\begin{split}
\boldsymbol{\psi}_{k}(n+1) &= \textbf{w}_{k}(n)+ \mu_{k} \hspace{0.3em}\textbf{U}^{T}_{k}(n)\left(\varepsilon I + \textbf{U}_{k}(n)\textbf{U}^{T}_{k}(n)\right)^{-1}  [\textbf{d}_{k}(n)- \textbf{U}^{T}_{k}(n)\textbf{w}_{k}(n)] \\
& \hspace{4em} + \mu_{k} \hspace{0.2em} \eta \sum\limits_{l \in \mathcal{N}_{k} \setminus \mathcal{C}(k)}{}  \hspace{0.2em} \rho^{'}_{kl}(n) \hspace{0.2em} \big[ \textbf{w}_{l}(n)-\textbf{w}_{k}(n) \big] \\
\textbf{w}_{k}(n+1) &= \sum\limits_{l \in \mathcal{N}_{k} \cap \mathcal{C}(k)}{} a_{lk} \hspace{0.2em} \boldsymbol{\psi}_{l}(n+1)   \\
\end{split}
\end{equation}
where $\rho^{'}_{kl}(n)=\rho_{kl} \hspace{0.2em} \delta_{kl}(n)$. Therefore, using the above expressions, the global model of modified multi-task diffusion
APA is formulated as follows:
\begin{equation}\label{eq4.4.4}
\begin{split}
\textbf{w}(n+1)&= \boldsymbol{\mathcal{A}} \hspace{0.2em} \Big[ \textbf{w}(n)+ \textbf{D} \hspace{0.2em} \textbf{U}^{T}(n) \big[\varepsilon \textbf{I} + \textbf{U}(n) \textbf{U}^{T}(n) \big]^{-1} \hspace{0.5em} \textbf{e}(n) \hspace{0.2em} - \textbf{D} \hspace{0.2em} \boldsymbol{\eta} \hspace{0.2em} \boldsymbol{\mathcal{Q}}_{\delta}(n) \hspace{0.2em}\textbf{w}(n)\Big]
\end{split}
\end{equation}
where
\begin{equation}\label{eq4.4.5}
\begin{split}
\boldsymbol{\mathcal{Q}}_{\delta} (n)&= \textbf{D}_{\delta} (n)- \textbf{P}_{\delta} (n) \otimes \textbf{I}_{L}
\end{split}
\end{equation}
and
\begin{equation}\label{eq4.4.6}
\begin{split}
\textbf{D}_{\delta} (n)&=\text{diag}\{\delta_{1}(n)\textbf{I}_{L}(n), \delta_{2}(n)\textbf{I}_{L}(n), \cdots, \delta_{N}(n)\textbf{I}_{L}(n) \}\\
\end{split}
\end{equation}
with
\begin{equation}\label{eq4.4.7}
\begin{split}
\delta_{k}(n)= \sum\limits_{l \in\mathcal{N}_{k} \setminus \mathcal{C}_{k}}^{}  \rho^{'}_{kl}(n) = \sum\limits_{l \in \mathcal{N}_{k} \setminus \mathcal{C}_{k}}^{} \rho_{kl} \hspace{0.3em} \delta_{kl}(n)
\end{split}
\end{equation}
the matrix $\textbf{P}_{\delta}(n)= \textbf{P} \odot \boldsymbol{\delta}(n)$ ('$\odot$' indicates the Hadamard product) is the $N \times N$ asymmetric matrix that defines regularizer strength among the nodes with $\boldsymbol{\delta}_{k}(n)=1$ and $\textbf{P}_{\delta, kk}(n) = 1$ if $\mathcal{N}_{k} \setminus \mathcal{C}(k)$ is empty. The matrices $\boldsymbol{\mathcal{A}}$, $\textbf{D}$, $\textbf{U}(n)$ are as same as the matrices defined in the network model section. Now the objective is to study the performance behavior of the multi-task diffusion APA governed
by the form $\eqref{eq4.4.4}$.
\subsection{Mean Error Behavior Analysis}
By denoting $\widetilde{\textbf{w}}(n)= \textbf{w}^{\star}-\textbf{w}(n)$
the recursive update equation of global weight error vector can be written as
\begin{equation}\label{eq4.5.1}
\begin{split}
\widetilde{\textbf{w}}(n+1)&= \boldsymbol{\mathcal{A}} \hspace{0.2em} \Big[ \textbf{I}_{LN} - \textbf{D} \hspace{0.2em} \textbf{Z}(n) \hspace{0.2em} - \hspace{0.2em} \textbf{D} \hspace{0.2em}\boldsymbol{\eta} \hspace{0.2em} \boldsymbol{\mathcal{Q}}_{\delta}(n) \Big]\widetilde{\textbf{w}}(n) - \boldsymbol{\mathcal{A}} \hspace{0.1em} \textbf{D} \hspace{0.1em} \textbf{U}^{T}(n) \big[\varepsilon \textbf{I} + \textbf{U}(n) \textbf{U}^{T}(n) \big]^{-1} \hspace{0.1em} \textbf{\emph{v}}(n) + \hspace{1em} \boldsymbol{\mathcal{A}} \hspace{0.2em} \textbf{D} \hspace{0.2em}\boldsymbol{\eta} \hspace{0.2em} \boldsymbol{\mathcal{Q}}_{\delta}(n) \hspace{0.2em} \textbf{w}^{\star} \\
\end{split}
\end{equation}
Taking the expectation of both sides, in addition to the statistical independence between $\textbf{w}_{k}(n)$ and $\textbf{U}_{k}(n)$ (i.e., "independence assumption") we are assuming statistical independence between $\textbf{w}_{k}(n)$ and $\delta_{kl}(n)$ and also recalling that $\textbf{\emph{v}}_{k}(n)$ is zero-mean
i.i.d and also independent of $\textbf{U}_{k}(n)$ and thus of $\textbf{w}_{k}(n)$ we can write
\begin{equation}\label{eq4.5.2}
\begin{split}
E\big[\widetilde{\textbf{w}}(n+1)\big]= \boldsymbol{\mathcal{A}} \bigg[ \textbf{I}_{LN} - \textbf{D} \hspace{0.2em}  \overline{\textbf{Z}} - \hspace{0.2em} \textbf{D} \hspace{0.2em} \boldsymbol{\eta} \hspace{0.2em} \overline{\boldsymbol{\mathcal{Q}}}_{\delta} \bigg] E[\widetilde{\textbf{w}}(n)] +  \boldsymbol{\mathcal{A}} \hspace{0.2em} \textbf{D} \hspace{0.2em} \boldsymbol{\eta} \hspace{0.2em} \overline{\boldsymbol{\mathcal{Q}}}_{\delta} \hspace{0.2em} \textbf{w}^{\star}
\end{split}
\end{equation}
where $\overline{\textbf{Z}}=E[\textbf{Z}(n)]$ and $\overline{\boldsymbol{\mathcal{Q}}}_{\delta}=E[\boldsymbol{\mathcal{Q}}_{\delta}(n)]$. The quantity $\overline{\boldsymbol{\mathcal{Q}}}_{\delta}$ is given as follows:
\begin{equation}\label{eq4.5.3}
\begin{split}
\overline{\boldsymbol{\mathcal{Q}}}_{\delta} = E\big[ \textbf{D}_{\delta} (n)- \textbf{P}_{\delta} (n) \otimes \textbf{I}_{L} \big] =  \overline{\textbf{D}}_{\delta} - \overline{\textbf{P}}_{\delta} \otimes \textbf{I}_{L}
\end{split}
\end{equation}
where  $\overline{\textbf{D}}_{\boldsymbol{\delta}}=E[\textbf{D}_{\boldsymbol{\delta}}]$, $\overline{\textbf{P}}_{\delta}=E[\textbf{P}_{\delta} (n)]$ and
\begin{equation}\label{eq4.5.3}
\begin{split}
\overline{\boldsymbol{\mathcal{Q}}}_{\delta, ij} = \begin{cases}
&\sum\limits_{l \in \mathcal{N}_{i} \setminus \mathcal{C}_{i}}^{} \rho_{i \hspace{0.1em}l} \hspace{0.2em}  E[\delta_{i \hspace{0.1em}l} (n) ]   \hspace{2em} \text{if}  \hspace{2em} i=j \\
& \rho_{i \hspace{0.1em}j} \hspace{0.2em}  E[\delta_{i \hspace{0.1em}j} (n) ]  \hspace{5em} \text{if}  \hspace{2em} i\neq j \\
& 0 \hspace{5em} \text{otherwise}
\end{cases}
\end{split}
\end{equation}
Then, for any initial condition, in order to guarantee the stability of the modified multi-task diffusion APA strategy in the mean sense, if the step size chosen to satisfy
\begin{equation}\label{eq4.5.3}
\begin{split}
 \lambda_{max} \Big( \boldsymbol{\mathcal{A}} \big[\textbf{I}_{LN} - \textbf{D} \hspace{0.2em} \overline{\textbf{Z}} - \textbf{D} \hspace{0.2em} \boldsymbol{\eta} \hspace{0.2em} \overline{\boldsymbol{\mathcal{Q}}}_{\delta}\big]\Big) < 1
\end{split}
\end{equation}
Now using the same arguments that are used in II. B, we will have
\begin{equation}\label{eq4.5.4}
\begin{split}
 \lambda_{max} \Big( \boldsymbol{\mathcal{A}} \big[\textbf{I}_{LN}-\textbf{D} \hspace{0.2em} \overline{\textbf{Z}} - \textbf{D} \hspace{0.2em} \boldsymbol{\eta} \hspace{0.2em} \overline{\boldsymbol{\mathcal{Q}}}_{\delta} \big] \Big) \leq \| \big[\textbf{I}_{LN}-\textbf{D} \hspace{0.2em} \overline{\textbf{Z}}-\textbf{D} \hspace{0.2em}\boldsymbol{\eta} \hspace{0.2em} \overline{\textbf{D}}_{\boldsymbol{\delta}}+\textbf{D} \hspace{0.2em}\boldsymbol{\eta} \hspace{0.2em} \big( \overline{\textbf{P}}_{\delta} \otimes \textbf{I}_{L}\big) \big] \|_{b, \infty}
\end{split}
\end{equation}
From Gershgorin circle theorem, a sufficient condition for $\eqref{eq4.5.3}$ to hold is to choose $\mu_{k}$ such that
\begin{equation}\label{eq4.5.6}
\begin{split}
0 <\mu_{k} < \frac{2}{\text{max}_{k}\{\lambda_{max}( \overline{\textbf{Z}}_{k})\}+ 2 \hspace{0.2em}\eta \hspace{0.2em} \text{max}_{k} (\overline{\boldsymbol{\delta}}_{k})}
\end{split}
\end{equation}
where $\overline{\boldsymbol{\delta}}_{k}= E[\boldsymbol{\delta}_{k} (n)] =\sum\limits_{l \in \mathcal{N}_{k} \setminus \mathcal{C}_{k}}^{} \rho_{kl} \hspace{0.2em} E[\delta_{kl} (n) ] $. Recalling the fact that $\delta_{kl}(n) $ is equal to either $0$ or $1$, we can write $0 \leq E[\delta_{kl} (n)] \leq 1$ that imply $\overline{\boldsymbol{\delta}}_{k} \leq 1$. Therefore, the presence of similarity measure, $\delta_{kl}(n)$ makes the modified multi-task diffusion strategy mean stability is better than the multi task diffusion strategy mentioned in $\eqref{eq3.10}$ however, lower than the diffusion APA due to the presence of $\eta$.
\par
In steady-state i.e., as $n \rightarrow \infty $  the asymptotic mean bias is given by
\begin{equation}\label{eq4.5.7}
\begin{split}
\lim\limits_{n \to \infty}E\big[\widetilde{\textbf{w}}(n)\big]= \Bigg[ \textbf{I}_{LN} -  \boldsymbol{\mathcal{A}} \bigg[ \textbf{I}_{LN} - \textbf{D} \hspace{0.2em} \overline{\textbf{Z}} \hspace{0.2em} - \hspace{0.2em} \textbf{D} \hspace{0.2em}\boldsymbol{\eta} \hspace{0.2em} \overline{\boldsymbol{\mathcal{Q}}}_{\delta} \bigg]  \Bigg]^{-1}  \boldsymbol{\mathcal{A}} \hspace{0.2em} \textbf{D} \hspace{0.2em} \boldsymbol{\eta} \hspace{0.2em} \overline{\boldsymbol{\mathcal{Q}}}_{\delta} \hspace{0.2em} \textbf{w}^{\star}
\end{split}
\end{equation}
\subsection{Mean-Square Error Behavior Analysis}
The recursive update equation of the modified multi-task diffusion APA weight error vector can also be rewritten as
\begin{equation}\label{eq4.6.1}
\begin{split}
\widetilde{\textbf{w}}(n+1)&= \boldsymbol{\mathcal{G}}_{\delta}(n) \widetilde{\textbf{w}}(n) - \boldsymbol{\mathcal{A}} \hspace{0.2em} \textbf{D} \hspace{0.2em} \textbf{U}^{T}(n) \big[\varepsilon \textbf{I} + \textbf{U}(n) \textbf{U}^{T}(n) \big]^{-1} \textbf{\emph{v}}(n) + \textbf{r}_{\delta}
\end{split}
\end{equation}
where
\begin{equation}\label{eq4.6.2}
\begin{split}
\boldsymbol{\mathcal{G}}_{\delta}(n)&= \boldsymbol{\mathcal{A}} \hspace{0.2em} \Big[\textbf{I}_{LN} -  \textbf{D} \hspace{0.2em} \textbf{Z}(n) - \textbf{D}\hspace{0.1em}\boldsymbol{\eta} \hspace{0.1em}\textbf{Q}_{  \delta}\Big]\\
\textbf{r}_{\delta}&= \boldsymbol{\mathcal{A}} \hspace{0.2em} \textbf{D} \hspace{0.2em} \boldsymbol{\eta} \hspace{0.2em} \boldsymbol{\mathcal{Q}}_{ \delta} \hspace{0.2em} \textbf{w}^{\star}
\end{split}
\end{equation}
In addition to standard independent assumption between $\textbf{U}_{k}(n)$ and $\textbf{w}_{k}(n)$ and  $E[\textbf{v}(n)]=0$ that was taken in II.C, here we assume statistical independence between $\delta_{kl}(n)$  and $\textbf{w}_{k}(n)$. Then the mean square of the weight error vector $\widetilde{\textbf{w}}(n+1)$, weighted by any positive semi-definite matrix $\Sigma$  that we are free to choose, satisfies the following relation:
\begin{equation}\label{eq4.6.3}
\begin{split}
E\|\widetilde{\textbf{w}}(n+1)\|_{\boldsymbol{\Sigma}}^{2}&= E\|\widetilde{\textbf{w}}(n)\|_{E \boldsymbol{\Sigma}^{'}_{\delta}}^{2}
+ E\big[\textbf{\emph{v}}^{T}(n) \hspace{0.1em}\textbf{Y}^{\boldsymbol{\Sigma}}(n) \hspace{0.1em} \textbf{\emph{v}}(n) \big]+ \hspace{0.2em}  E\big[\widetilde{\textbf{w}}(n)\big]^{T} E\big[\boldsymbol{\mathcal{G}}_{\delta}(n) \big]^{T} \boldsymbol{\Sigma} \hspace{0.2em} \textbf{r}_{\delta} + \textbf{r}_{\delta}^{T} \boldsymbol{\Sigma} \hspace{0.2em} E\big[\boldsymbol{\mathcal{G}}_{\delta}(n)\big]\hspace{0.1em} E\big[\widetilde{\textbf{w}}(n)\big] + \|\textbf{r}_{\delta}\|_{\boldsymbol{\Sigma}}^{2}
\end{split}
\end{equation}
where
\begin{equation}\label{eq4.6.4}
\begin{split}
E\boldsymbol{\Sigma}^{'}_{\delta}&= E \Big[\boldsymbol{\mathcal{G}}_{\delta}^{T}(n) \hspace{0.2em} \boldsymbol{\Sigma} \hspace{0.2em} \boldsymbol{\mathcal{G}}_{\delta}(n) \Big]\\
&=\hspace{0.2em} \boldsymbol{\mathcal{A}}^{T} \hspace{0.1em} \boldsymbol{\Sigma}\hspace{0.1em} \boldsymbol{\mathcal{A}} -  \overline{\textbf{Z}} \hspace{0.2em} \textbf{D} \hspace{0.2em} \boldsymbol{\mathcal{A}}^{T} \hspace{0.1em} \boldsymbol{\Sigma} \hspace{0.2em} \boldsymbol{\mathcal{A}} - \boldsymbol{\mathcal{A}}^{T} \hspace{0.2em} \boldsymbol{\Sigma} \hspace{0.2em} \boldsymbol{\mathcal{A}} \hspace{0.2em} \textbf{D} \hspace{0.2em}   \overline{\textbf{Z}} - \boldsymbol{\mathcal{A}}^{T} \hspace{0.2em} \boldsymbol{\Sigma} \hspace{0.2em} E[\boldsymbol{\mathcal{S}}_{\delta}] - E[\boldsymbol{\mathcal{S}}_{\delta}^{T}] \hspace{0.2em} \boldsymbol{\Sigma} \hspace{0.2em} \boldsymbol{\mathcal{A}} \hspace{0.1em}\\
&\hspace{0.2em} +  \overline{\textbf{Z}} \hspace{0.2em} \textbf{D} \hspace{0.2em} \boldsymbol{\mathcal{A}}^{T} \hspace{0.2em} \boldsymbol{\Sigma} \hspace{0.2em} E[\boldsymbol{\mathcal{S}}_{\delta}] + E[\boldsymbol{\mathcal{S}}_{\delta}^{T}] \hspace{0.2em} \boldsymbol{\Sigma} \hspace{0.2em} \boldsymbol{\mathcal{A}} \hspace{0.2em} \textbf{D} \hspace{0.2em} \overline{ \textbf{Z}} + E \big[\textbf{U}^{T}(n)\hspace{0.1em} \textbf{Y}^{\boldsymbol{\Sigma}}\hspace{0.1em} \textbf{U}(n) \big] + E[\boldsymbol{\mathcal{S}}_{\delta}^{T} \hspace{0.2em} \boldsymbol{\Sigma} \hspace{0.2em} \boldsymbol{\mathcal{S}}_{\delta}]
\end{split}
\end{equation}
and
\begin{equation}\label{eq4.6.5}
\begin{split}
\textbf{Y}^{\boldsymbol{\Sigma}}&=\big[\varepsilon \textbf{I} + \textbf{U}(n) \textbf{U}^{T}(n) \big]^{-1}\textbf{U}(n) \hspace{0.2em} \textbf{D} \hspace{0.2em} \boldsymbol{\mathcal{A}}^{T} \hspace{0.2em} \boldsymbol{\Sigma} \hspace{0.2em} \boldsymbol{\mathcal{A}} \hspace{0.2em}\textbf{D} \hspace{0.2em} \textbf{U}^{T}(n) \big[\varepsilon \textbf{I} + \textbf{U}(n) \textbf{U}^{T}(n) \big]^{-1}\\
\boldsymbol{\mathcal{S}}_{\delta}&= \boldsymbol{\mathcal{A}} \hspace{0.2em} \textbf{D} \hspace{0.2em} \boldsymbol{\eta} \hspace{0.2em} \boldsymbol{\mathcal{Q}}_{\delta}
\end{split}
\end{equation}
Following the same procudre mentioned in II. C, to extract the matrix $\boldsymbol{\Sigma}$ from the expectation terms, a weighted variance relation is introduced by using $L^{2}N^{2} \times 1$ column vectors:
\begin{equation}\label{eq4.6.7}
\begin{split}
\boldsymbol{\sigma}= \text{bvec}\{\boldsymbol{\Sigma}\} \hspace{2em} \text{and} \hspace{2em} \boldsymbol{\sigma}_{\delta}= \text{bvec}\{E\boldsymbol{\Sigma}_{\delta}\}
\end{split}
\end{equation}
 with a linear relation between the corresponding vectors $\{\boldsymbol{\sigma}, \boldsymbol{\sigma}_{\delta}\}$
\begin{equation}\label{eq4.6.18}
\begin{split}
\boldsymbol{\sigma}_{\delta}= \textbf{F}_{\delta} \hspace{0.2em} \boldsymbol{\sigma}
\end{split}
\end{equation}
where $\textbf{F}_{\delta}$ is an $L^{2}N^{2} \times L^{2}N^{2}$ matrix and given by
\begin{equation}\label{eq4.6.19}
\begin{split}
\textbf{F}_{\delta}&=\left[\begin{array}{l}  \textbf{I}_{LN}- \big(\textbf{I}_{LN} \otimes_{b} \overline{\textbf{Z}} \big)  \big(\textbf{I}_{LN} \otimes_{b} \textbf{D} \big)  - \big(  \overline{\textbf{Z}} \otimes_{b} \textbf{I}_{LN}\big)  \big(\textbf{D} \otimes_{b} \textbf{I}_{LN}\big) \\
 - \hspace{0.5em} \big( \overline{\boldsymbol{\mathcal{Q}}}_{\delta}^{T} \otimes_{b} \textbf{I}_{LN}\big) \big(\boldsymbol{\eta}  \otimes_{b} \textbf{I}_{LN}\big) \big(\textbf{D}  \otimes_{b} \textbf{I}_{LN}\big) -  \big(\textbf{I}_{LN} \otimes_{b}  \overline{\boldsymbol{\mathcal{Q}}}_{\delta}^{T}  \big) \big(\textbf{I}_{LN} \otimes_{b} \boldsymbol{\eta}  \big) \big(\textbf{I}_{LN} \otimes_{b} \textbf{D}  \big) \\
 + \hspace{0.5em} \big( \textbf{I}_{LN} \otimes_{b}  \overline{\textbf{Z}}   \big) \big( \overline{\boldsymbol{\mathcal{Q}}}^{T}_{\delta}  \otimes_{b} \textbf{I}_{LN}  \big) \big(\boldsymbol{\eta}  \otimes_{b} \textbf{I}_{LN}\big) \big(\textbf{D} \otimes_{b} \textbf{D} \big)\\
+ \hspace{0.5em} \big( \overline{\textbf{Z}} \otimes_{b} \textbf{I}_{LN}\big) \big(\textbf{I}_{LN} \otimes_{b} \overline{\boldsymbol{\mathcal{Q}}}_{\delta}^{T}  \big) \big(\textbf{I}_{LN} \otimes_{b} \boldsymbol{\eta}  \big) \big(\textbf{D}  \otimes_{b} \textbf{D}  \big) +  \boldsymbol{\Pi} \big(\textbf{D} \otimes_{b} \textbf{D}  \big)\\
+  \hspace{0.5em} \big( E[ \boldsymbol{\mathcal{Q}}^{T}_{\delta} \otimes_{b} \boldsymbol{\mathcal{Q}}^{T}_{\delta}]  \big) \big(\boldsymbol{\eta}  \otimes_{b} \boldsymbol{\eta}  \big) \big(\textbf{D} \otimes_{b} \textbf{D}  \big)   \end{array} \right] \big( \boldsymbol{\mathcal{A}}^{T}  \odot \boldsymbol{\mathcal{A}}^{T} \big)
\end{split}
\end{equation}
where $\boldsymbol{\Pi}=E\big[\textbf{Z}(n) \otimes_{b} \textbf{Z}(n)  \big]$
\par
The noise term $E\big[\textbf{\emph{v}}^{T}(n) \hspace{0.1em}\textbf{Y}^{\boldsymbol{\Sigma}}(n) \hspace{0.1em} \textbf{\emph{v}}(n) \big] = \boldsymbol{\gamma}^{T} \hspace{0.1em} \boldsymbol{\sigma}$.
\par
Finally, let us define the $\emph{\textbf{f}}\big(\textbf{r}_{\delta}, E[\widetilde{\textbf{w}}(n)], \boldsymbol{\sigma}\big)$ as the last three terms on the right hand side of the $\eqref{eq4.3.3}$, i.e,
\begin{equation}\label{eq4.6.23}
\begin{split}
\emph{\textbf{f}}\big(\textbf{r}_{\delta}, E[\widetilde{\textbf{w}}(n)], \boldsymbol{\sigma}\big)&=   \|\textbf{r}_{\delta}\|_{\boldsymbol{\Sigma}}^{2} + \hspace{0.2em}  E\big[\widetilde{\textbf{w}}(n)\big]^{T} E\big[\boldsymbol{\mathcal{G}}_{\delta}(n) \big]^{T}  \boldsymbol{\Sigma} \hspace{0.2em} \textbf{r}_{\delta} + \textbf{r}_{\delta}^{T} \boldsymbol{\Sigma} \hspace{0.2em} E\big[\textbf{G}_{\delta}(n)\big]\hspace{0.1em} E\big[\widetilde{\textbf{w}}(n)\big] \\
&= \Big(\textbf{r}^{T}_{b, \delta} + \boldsymbol{\alpha}^{T}_{\delta, 1} + \boldsymbol{\alpha}^{T}_{\delta, 2} \big) \hspace{0.2em} \boldsymbol{\sigma}
\end{split}
\end{equation}
where
\begin{equation}
\begin{split}
\textbf{r}_{b, \delta}&= E\Big( \boldsymbol{\mathcal{A}}  \otimes_{b} \boldsymbol{\mathcal{A}}  \Big) \Big( \textbf{D} \otimes_{b} \textbf{D} \Big) \Big( \boldsymbol{\eta} \otimes_{b} \boldsymbol{\eta} \Big) \Big( E\Big[\boldsymbol{\mathcal{Q}}_{\delta}  \otimes_{b} \boldsymbol{\mathcal{Q}}_{\delta} \big] \Big)  \text{bvec}\Big\{ \textbf{w}^{\star} \big(\textbf{w}^{\star}\big)^{T} \Big\}\\
\end{split}
\end{equation}
\begin{equation}
\begin{split}
\boldsymbol{\alpha}_{\delta, 1}(n)&=  \Big( \boldsymbol{\mathcal{A}}  \otimes_{b} \boldsymbol{\mathcal{A}}  \Big) \left[\begin{array}{l}  \big( \textbf{I}_{LN} \otimes_{b}  \textbf{D}  \big)  \big( \textbf{I}_{LN} \otimes_{b} \boldsymbol{\eta} \big) \Big( \textbf{I}_{LN} \otimes_{b} \overline{\boldsymbol{\mathcal{Q}}}_{\delta}   \Big) \\
-  \big( \textbf{D}  \otimes_{b} \textbf{D}  \big)  \big( \textbf{I}_{LN} \otimes_{b} \boldsymbol{\eta} \big) \Big( \textbf{I}_{LN} \otimes_{b} \overline{\boldsymbol{\mathcal{Q}}}_{\delta}   \Big) \big( \overline{\textbf{Z}} \otimes_{b} \textbf{I}_{LN} \big)  \\
-  \big( \textbf{D}  \otimes_{b} \textbf{D}  \big) \big( \boldsymbol{\eta} \otimes_{b} \boldsymbol{\eta} \big)  \Big( \big[\boldsymbol{\mathcal{Q}}_{\delta}  \otimes_{b} \boldsymbol{\mathcal{Q}}_{\delta} \big]  \Big) \end{array}\right] \text{bvec}\Big\{\textbf{w}^{\star} E[\widetilde{\textbf{w}}^{T}(n) ] \Big\}\\
\end{split}
\end{equation}
and
\begin{equation}
\begin{split}
\boldsymbol{\alpha}_{\delta, 2}(n)&=  \Big( \boldsymbol{\mathcal{A}}  \otimes_{b} \boldsymbol{\mathcal{A}}  \Big) \left[\begin{array}{l}  \big( \textbf{D}  \otimes_{b} \textbf{I}_{LN} \big)  \big( \boldsymbol{\eta} \otimes_{b} \textbf{I}_{LN} \big) \Big( \overline{\boldsymbol{\mathcal{Q}}}_{\delta}  \otimes_{b} \textbf{I}_{LN} \Big) \\
-  \big( \textbf{D}  \otimes_{b} \textbf{D}  \big)  \big( \boldsymbol{\eta} \otimes_{b} \textbf{I}_{LN} \big) \Big(  \overline{\boldsymbol{\mathcal{Q}}}_{\delta}  \otimes_{b} \textbf{I}_{LN} \Big)  \big( \textbf{I}_{LN} \otimes_{b} \overline{\textbf{Z}} \big)  \\
- \big( \textbf{D}  \otimes_{b} \textbf{D}  \big) \big( \boldsymbol{\eta} \otimes_{b} \boldsymbol{\eta} \big)  \Big( E\big[ \boldsymbol{\mathcal{Q}}_{\delta}  \otimes_{b} \boldsymbol{\mathcal{Q}}_{\delta} \big]  \Big) \end{array}\right] \text{bvec}\Big\{ E[\widetilde{\textbf{w}}(n)] \big(\textbf{w}^{\star} \big)^{T}  \Big\}\\
\end{split}
\end{equation}
Therefore, the mean-square behavior of the modified multi-task diffusion APA algorithm is summarized
as follows:
\begin{equation}\label{eq4.6.24}
\begin{split}
E\|\widetilde{\textbf{w}}(n+1)\|_{\boldsymbol{\sigma}}^{2}&= E\|\widetilde{\textbf{w}}(n)\|_{\textbf{F}_{\delta} \hspace{0.2em} \boldsymbol{\sigma}}^{2}
+ \boldsymbol{\gamma}^{T} \hspace{0.1em} \boldsymbol{\sigma}+\emph{\textbf{f}}\big(\textbf{r}_{\delta}, E[\widetilde{\textbf{w}}(n)], \boldsymbol{\sigma}\big)
\end{split}
\end{equation}
Therefore, the modified multi-task diffusion APA strategy presented in $\eqref{eq4.4.3}$ is mean square stable if the matrix $\textbf{F}_{\delta}$ is stable. Iterating the recursion $\eqref{eq4.6.24}$ starting from $n=0$, we get
\begin{equation}\label{eq4.6.25}
\begin{split}
E\|\widetilde{\textbf{w}}(n+1)\|_{\boldsymbol{\sigma}}^{2}&= E\|\widetilde{\textbf{w}}(0)\|_{\textbf{F}_{\delta}^{n+1}\boldsymbol{\sigma}}^{2}
+ \boldsymbol{\gamma}^{T} \hspace{0.1em} \sum\limits_{i=0}^{n}\textbf{F}_{\delta}^{i}\boldsymbol{\sigma}+\sum\limits_{i=0}^{n}\emph{\textbf{f}}\big(\textbf{r}_{\delta}, E[\widetilde{\textbf{w}}(n-i)], \textbf{F}_{\delta}^{i} \boldsymbol{\sigma}\big)
\end{split}
\end{equation}
with initial condition $\widetilde{\textbf{w}}(0)=\textbf{w}^{\star} - \textbf{w}(0)$. If the matrix $\textbf{F}_{\delta}$ is stable then the first and second terms in the above equation converge to a finite value as $n \rightarrow \infty $. Now, let us consider the third term on the RHS of the $\eqref{eq4.6.25}$. We know that $E[\widetilde{\textbf{w}}(n)]$ is uniformly bounded because $\eqref{eq4.5.2}$ is a BIBO stable recursion with bounded driving term $ \hspace{0.1em} \boldsymbol{\mathcal{A}} \hspace{0.2em}\textbf{D} \hspace{0.2em} \boldsymbol{\eta} \hspace{0.2em} E[\boldsymbol{\mathcal{Q}}_{\delta}] \hspace{0.2em} \textbf{w}^{\star}$. Therefore, from $\eqref{eq4.6.23}$  $\emph{\textbf{f}}\big(\textbf{r}_{\delta}, E[\widetilde{\textbf{w}}(n-i)], \textbf{F}_{\delta}^{i} \boldsymbol{\sigma}\big)$ can be written as
\begin{equation}\label{eq4.6.26}
\begin{split}
\emph{\textbf{f}}\big(\textbf{r}_{\delta}, E[\widetilde{\textbf{w}}(n-i)], \textbf{F}_{\delta}^{i} \boldsymbol{\sigma}\big)&= \Big(\textbf{r}^{T}_{b, \delta}  +  \boldsymbol{\alpha}^{T}_{\delta, 1} (n-i) +  \boldsymbol{\alpha}^{T}_{\delta, 2} (n-i) \Big)  \hspace{0.2em} \textbf{F}_{\delta}^{i}  \hspace{0.2em} \boldsymbol{\sigma}
\end{split}
\end{equation}
Provided that $\textbf{F}_{\delta}$ is stable and there exist a matrix norm, denoted by $\|\cdot\|_{p}$ such that $\|\textbf{F}_{\delta}\|_{p}=c_{p, \delta } < 1$. Applying this norm to $\emph{\textbf{f}}$ and using the matrix norms and triangular inequality, we can write $ \| \emph{\textbf{f}}\big(\textbf{r}_{\delta}, E[\widetilde{\textbf{w}}(n-i)], \textbf{F}_{\delta}^{i} \hspace{0.2em} \boldsymbol{\sigma}\big)  \| \leq \emph{v} \hspace{0.1em} c^{i}_{p}$, given $\emph{v}$ is a small positive constant. Therefore $E\|\widetilde{\textbf{w}}(n+1)\|_{\boldsymbol{\sigma}}^{2}$ converges to a bounded value as $n \rightarrow \infty$, and the algorithm is said to be mean square stable.
\par
By selecting $\boldsymbol{\Sigma}=\frac{1}{N} \textbf{I}_{LN}$ we can relate $E\|\widetilde{\textbf{w}}(n+1)\|_{\boldsymbol{\sigma}}^{2}$ and $E\|\widetilde{\textbf{w}}(n)\|_{\boldsymbol{\sigma}}^{2}$ as follows:
\begin{equation}\label{eq4.6.27}
\begin{split}
E\|\widetilde{\textbf{w}}(n+1)\|_{\boldsymbol{\sigma}}^{2}&=E\|\widetilde{\textbf{w}}(n)\|_{\boldsymbol{\sigma}}^{2} + \boldsymbol{\gamma}^{T} \textbf{F}_{\delta}^{n} \hspace{0.2em} \boldsymbol{\sigma} - E\|\widetilde{\textbf{w}}(0)\|_{\big(I_{(LN)^{2}}  - \textbf{F}_{\delta} \hspace{0.1em}\big) \textbf{F}_{\delta}^{n} \hspace{0.2em} \boldsymbol{\sigma}}^{2} \\
& \hspace{2em} + \sum\limits_{i=0}^{n}\emph{\textbf{f}}\big(\textbf{r}_{\delta}, E[\widetilde{\textbf{w}}(n-i)], \textbf{F}_{\delta}^{i} \hspace{0.2em} \boldsymbol{\sigma}\big)  - \sum\limits_{i=0}^{n-1}\emph{\textbf{f}}\big(\textbf{r}_{\delta}, E[\widetilde{\textbf{w}}(n-1-i)], \textbf{F}_{\delta}^{i} \hspace{0.2em} \boldsymbol{\sigma}\big)
\end{split}
\end{equation}
we can rewrite the last two terms in the above equation as,
\begin{equation}\label{eq4.6.28}
\begin{split}
&\sum\limits_{i=0}^{n}\emph{\textbf{f}}\big(\textbf{r}_{\delta}, E[\widetilde{\textbf{w}}(n-i)], \textbf{F}_{\delta}^{i} \hspace{0.2em} \boldsymbol{\sigma}\big) - \sum\limits_{i=0}^{n-1}\emph{\textbf{f}}\big(\textbf{r}_{\delta}, E[\widetilde{\textbf{w}}(n-1-i)], \textbf{F}_{\delta}^{i} \hspace{0.2em} \boldsymbol{\sigma}\big)  = \textbf{r}_{b, \delta}^{T} \hspace{0.2em} \textbf{F}^{n} \hspace{0.2em} \boldsymbol{\sigma} + \left[   \boldsymbol{\alpha}_{\delta, 1}^{T}(n) + \boldsymbol{\alpha}_{\delta, 2}^{T}(n)   + \boldsymbol{\Gamma}_{\delta}(n)  \right] \hspace{0.2em}  \boldsymbol{\sigma}
\end{split}
\end{equation}
where
\begin{equation}\label{eq4.6.29}
\begin{split}
\boldsymbol{\Gamma_{\delta}} (n) = \sum\limits_{i=1}^{n} \Big( \boldsymbol{\alpha}_{1, \delta}^{T}(n-i) +  \boldsymbol{\alpha}_{2, \delta}^{T}(n-i) \Big) \textbf{F}_{\delta}^{i} \hspace{0.2em} \boldsymbol{\sigma} - \sum\limits_{i=0}^{n-1} \Big( \boldsymbol{\alpha}_{1, \delta}^{T}(n-1-i) +  \boldsymbol{\alpha}_{2, \delta}^{T}(n-1-i) \Big) \textbf{F}_{\delta}^{i} \hspace{0.2em} \boldsymbol{\sigma}
\end{split}
\end{equation}
Therefore, the recursion presented in $\eqref{eq4.6.24}$ can be rewritten as,
\begin{equation}\label{eq4.6.30}
\begin{split}
E\|\widetilde{\textbf{w}}(n+1)\|_{\boldsymbol{\sigma}}^{2}&= E\|\widetilde{\textbf{w}}(n)\|_{\boldsymbol{\sigma}}^{2} + \boldsymbol{\gamma}^{T} \textbf{F}_{\delta}^{n} \boldsymbol{\sigma} - E\|\widetilde{\textbf{w}}(0)\|_{\big(I_{(LN)^{2}} - \textbf{F}_{\delta} \hspace{0.1em}\big)\textbf{F}_{\delta}^{n} \boldsymbol{\sigma}}^{2} +\textbf{r}_{b, \delta}^{T} \hspace{0.2em} \textbf{F}_{\delta}^{n} \hspace{0.1em} \boldsymbol{\sigma} + \left[   \boldsymbol{\alpha}_{1, \delta}^{T}(n) + \boldsymbol{\alpha}_{2, \delta}^{T}(n)   + \boldsymbol{\Gamma_{\delta}}(n)  \right] \\
\boldsymbol{\Gamma_{\delta}} (n+1)&= \boldsymbol{\Gamma_{\delta}} (n) \textbf{F} + \Big[  \big[ \boldsymbol{\alpha}_{1, \delta}^{T}(n) + \boldsymbol{\alpha}_{2, \delta}^{T}(n)  \big]\hspace{0.2em} [\textbf{F}_{\delta} - \textbf{I}_{(LN)^{2}}]  \Big]
\end{split}
\end{equation}
with $\boldsymbol{\Gamma_{\delta}} (0)= 0_{1 \times (LN)^{2}}$.
\par
Steady-state MSD of the modified multi-task diffusion APA strategy is given as follows
\begin{equation}\label{eq4.6.31}
\begin{split}
\lim\limits_{n \to \infty}E\|\widetilde{\textbf{w}}(n)\|_{\big(I_{(LN)^{2}}-\textbf{F}_{\delta}\big) \boldsymbol{\sigma}}^{2}&= \boldsymbol{\gamma}^{T} \boldsymbol{\sigma} + \emph{\textbf{f}}\big(\textbf{r}_{\delta}, E[\widetilde{\textbf{w}}(\infty)], \boldsymbol{\sigma}\big)
\end{split}
\end{equation}

\section{Simulation Results}
A network consists of $9$ nodes with the topology shown in Fig. 2 was considered for simulations. The nodes were divided into $3$ clusters: $\mathcal{C}_{1} = \{1, 2, 3\}, \mathcal{C}_{2} = \{4, 5, 6\},$ and $\mathcal{C}_{3} = \{7, 8, 9\}$. First, for theoretical performance comparison purpose, we first considered randomly generated two dimension vectors of the form $\textbf{w}^{*}_{\mathcal{C}_{k}}=\textbf{w}_{0}+ \delta_{\mathcal{C}_{k}} \textbf{w}_{\mathcal{C}_{k}}$ where $\delta_{\mathcal{C}_{1}}=0.025, \delta_{\mathcal{C}_{2}}=-0.025$ and $\delta_{\mathcal{C}_{3}}=0.015$.
\begin{figure}[h]
\centering
\includegraphics [height=65mm,width=80mm]{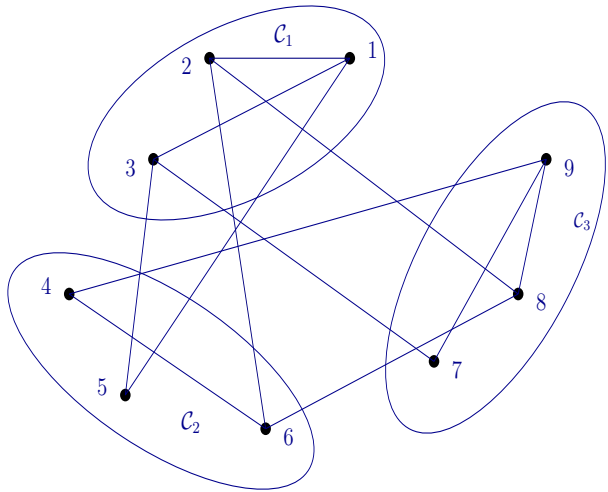}
\caption{Network Topology}
\label{the-label-for-cross-referencing}
\end{figure}
The input regressors $\textbf{u}_{k}(n)$ were taken from zero mean, Gaussian distribution with correlation matrices $\textbf{R}_{u, k}= I_{L}$ and the observation noises were i. i. d zero-mean Gaussian random variables, independent of any other signals with noise variance $\sigma^{2}_{v}=0.001$. The multi-task diffusion APA algorithm was run with different step sizes and regularization parameters. Regularization strength $\rho_{kl}$ was set to $\rho_{kl}= |\mathcal{N}_{k}\setminus\mathcal{C}(k)|^{-1}$ for $l \in \mathcal{N}_{k}\setminus\mathcal{C}(k)$, and $\rho_{kl}=0$ for any other $l$. This settings usually leads to asymmetrical regularization weights. The coefficient matrix $C$ was taken to be identity matrix and the combiner coefficients $a_{lk}$ were set according to Metropolis rule.
\par
Simulations were carried out to illustrate the performance of several learning strategies: $1)$ the non-cooperative APA algorithm, $2)$ the multi-task algorithm (Algorithm $3$), and $3)$ the clustered multi-task algorithm (Algorithm $1$). The non-cooperative algorithm was obtained by assigning a cluster to each node and setting $\eta=0$. The multi-task algorithm was obtained by assigning a cluster to each node and setting $\eta\neq0$. Note that algorithm $2$ was not considered for comparison since it is a single-task estimation method. Normalized MSD was taken as the performance parametric to compare the diffusion strategies. Projection order was taken to be $4$ and the initial taps were chosen to be zero.
\begin{figure}[h!]
\centering
\includegraphics [height=85mm,width=100mm]{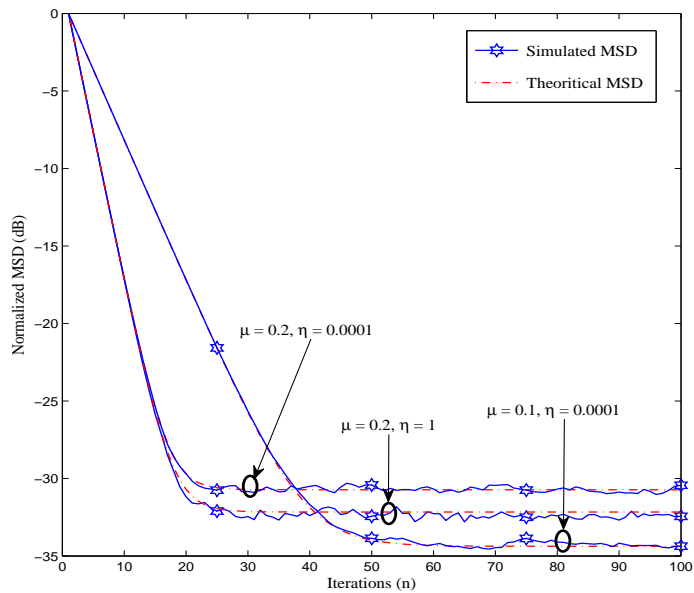}
\caption{Comparison of transient NMSD for different step-sizes and regularization parameters.}
\label{the-label-for-cross-referencing}
\end{figure}
\par
Secondly, the modified multi-task diffusion APA is compared with multi-task diffusion APA. For that, randomly generated coefficient vectors of the form $\textbf{w}^{*}_{\mathcal{C}_{k}}=\textbf{w}_{0}+ \delta_{\mathcal{C}_{k}} \textbf{w}_{\mathcal{C}_{k}}$ with $L=256$ taps length were chosen as $\delta_{\mathcal{C}_{1}}=0.025, \delta_{\mathcal{C}_{2}}=-0.025$ and $\delta_{\mathcal{C}_{3}}=0.015$. The input signal vectors were taken from zero mean, Gaussian distribution with correlation statistics as shown in the Fig. 4, and the observation noises were i. i. d zero-mean Gaussian random variables, independent of any other signals with noise variances as shown in the Fig. 5.
\begin{figure}[h!]
\centering
\includegraphics [height=35mm,width=100mm]{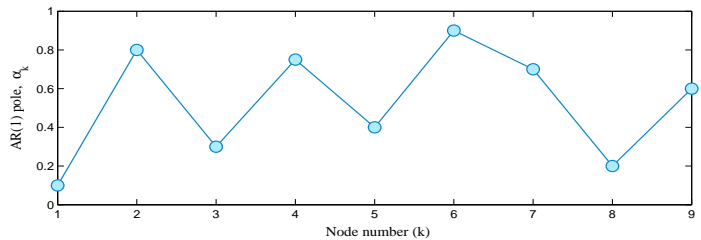}
\caption{Input signal Statistics.}
\label{the-label-for-cross-referencing}
\end{figure}

\begin{figure}[h!]
\centering
\includegraphics [height=35mm,width=100mm]{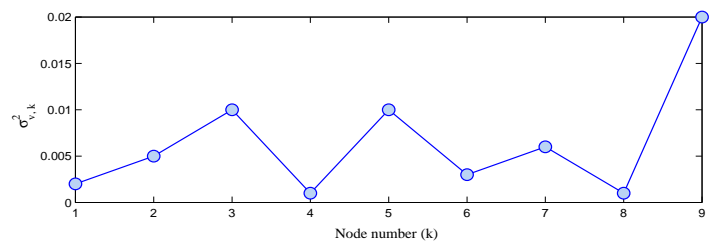}
\caption{Noise statistics.}
\label{the-label-for-cross-referencing}
\end{figure}
Projection order was taken to be $8$ and the initial taps were chosen to be zero. The step-size and regularization parameters $(\mu, \eta)$ were adjusted to compare the steady state MSD and convergence rate properly. Simulation results were obtained by averaging $50$ Monte-Carlo runs. The learning curves of diffusion strategies were presented in Fig. $6$. It can be observed that the performance of the non-cooperative strategy was poor as nodes do not collaborate for additional benefit. In the case of multi-task diffusion strategy the performance is improved over non-cooperative strategy due to regularization between nodes. The cluster information in addition to regularization among nodes in the clustered multi task results in better performance over the non-cooperative and multi task diffusion strategies. The extra information information in the regularization among nodes results in great improvement in the performance of modified clustered multi-task diffusion strategy.
\begin{figure}[h]
\centering
\includegraphics [height=110mm,width=120mm]{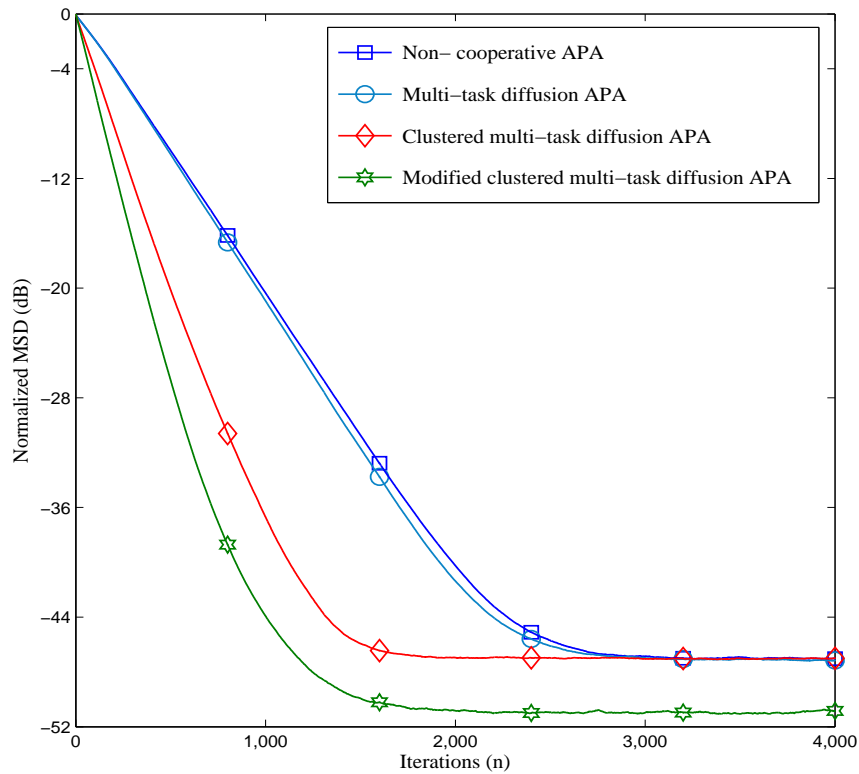}
\caption{Comparison of Modified Multi-task diffusion APA with multi-task diffusion APA}
\label{the-label-for-cross-referencing}
\end{figure}
\section{Conclusions}
In this paper, we presented the diffusion APA strategies which are suitable for multi-task networks and also robust against the correlated input conditions. The performance analysis of the proposed multi-task diffusion APA is presented in mean and mean square sense. By introducing similarity measure, the modified multi-task diffusion APA algorithms is proposed to achieve the improved performance over the multi-task diffusion strategies existed in literature.

\end{document}